\pdfoutput=1
\documentclass[sigconf,nonacm]{acmart}

\usepackage{amsmath,amssymb,amsfonts,amsthm,xspace}
\usepackage[linesnumbered,ruled,vlined]{algorithm2e}
\usepackage{algorithmic}
\usepackage{graphicx,subfigure,stfloats}
\usepackage{epstopdf}
\usepackage{textcomp}
\usepackage{ragged2e, xcolor}
\usepackage{balance}
\usepackage{verbatim}
\usepackage{enumitem}
\usepackage{makecell}
\usepackage{url}
\usepackage{booktabs}
\usepackage{multicol}
\usepackage{multirow}
\usepackage{todonotes}
\usepackage{diagbox}

\AtBeginDocument{%
  }

\acmConference[Conference acronym 'XX]{Make sure to enter the correct
  conference title from your rights confirmation emai}{June 03--05,
  2018}{Woodstock, NY}
\acmISBN{978-1-4503-XXXX-X/18/06}

\newcommand{\ignore}[1]{}
\newcommand{\nop}[1]{}
\newcommand{\eat}[1]{}
\newcommand{\kw}[1]{{\ensuremath{\mathsf{#1}}}\xspace}

\newcommand{\stitle}[1]{\vspace{1ex} \noindent{\bf #1}}
\long\def\comment#1{}

\newtheorem{definition}{Definition}
\newtheorem{theorem}{Theorem}

\newtheorem{example}{Example}

\newcommand{\Br}{\ensuremath{\mathcal{B}}\xspace}
\newcommand{\PPr}{\ensuremath{\mathcal{P}}\xspace}
\newcommand{\pbcpc}{\kw{PBCPC}}
\newcommand{\pbcpcp}{\kw{PBCPC+}}
\newcommand{\biclique}{\kw{Biclique}}
\newcommand{\bicliquem}{\kw{BicliqueM}}
\newcommand{\bicliquep}{\kw{BicliqueP}}

\newcommand{\showproof}{}

\begin{document}

\title{A Computationally Efficient Framework for Overlapping Community Detection in Large Bipartite Graphs}

\author{Yue Zeng}
\affiliation{%
	\institution{Beijing Institute of Technology}
	\country{}
}
\email{bruceez@163.com}

\author{Rong-Hua Li}
\affiliation{%
	\institution{Beijing Institute of Technology}
	\country{}
}
\email{lironghuabit@126.com}

\author{Qiangqiang Dai}
\affiliation{%
	\institution{Beijing Institute of Technology}
	\country{}
}
\email{qiangd66@gmail.com}

\author{Guoren Wang}
\affiliation{%
	\institution{Beijing Institute of Technology}
	\country{}
}
\email{wanggrbit@126.com}


\def\MBAG{\kw{MBAG}}
\def\BCPC{\kw{BCPC}}
\def\BCPCs{\kw{BCPC{\text{s}}}}
\begin{abstract}
Community detection, which uncovers closely connected vertex groups in networks, is vital for applications in social networks, recommendation systems, and beyond. Real-world networks often have bipartite structures (vertices in two disjoint sets with inter-set connections), creating unique challenges on specialized community detection methods. Biclique percolation community (\BCPC) is widely used to detect cohesive structures in bipartite graphs. A biclique is a complete bipartite subgraph, and a \BCPC forms when maximal bicliques connect via adjacency (sharing an \((\alpha, \beta)\)-biclique). Yet, existing methods for \BCPC detection suffer from high time complexity due to the potentially massive maximal biclique adjacency graph (\MBAG). To tackle this, we propose a novel partial-\BCPC based solution, whose key idea is to use partial-\BCPC to reduce the size of the \MBAG. A partial-\BCPC is a subset of \BCPC. Maximal bicliques belonging to the same partial-\BCPC must also belong to the same \BCPC. Therefore, these maximal bicliques can be grouped as a single vertex in the \MBAG, significantly reducing the size of the \MBAG. Furthermore, we move beyond the limitations of \MBAG and propose a novel \BCPC detection approach based on \((\alpha, \beta)\)-biclique enumeration. This approach detects \BCPC by enumerating all \((\alpha, \beta)\)-bicliques and connecting maximal bicliques sharing the same \((\alpha, \beta)\)-biclique, which is the condition for maximal bicliques to be adjacent. It also leverages partial-\BCPC to significantly prune the enumeration space of \((\alpha, \beta)\)-biclique. Experiments show that our methods outperform existing methods by nearly three orders of magnitude.
\end{abstract}



\maketitle

\section{Introduction}

Community detection aims to uncover groups of closely connected vertices in a network and is a core task in network analysis \cite{fortunato2010community}. It plays a key role in applications such as social networks \cite{steinhaeuser2008community}, biological systems \cite{rahiminejad2019topological} and recommendation systems \cite{choudhary2023community}. In many real-world scenarios, networks are naturally modeled as bipartite graphs, represented by $G=(U, V, E)$, where vertices are divided into two disjoint sets $U$ and $V$, and connections only occur between $ U $ and $ V $. Representative examples include user–item networks in recommendation systems \cite{maier2022bipartite} and author–paper networks in bibliometrics \cite{carusi2020look}. This structure introduces new challenges for community detection, motivating the development of specialized methods for bipartite graphs.

The \underline{b}i\underline{c}lique \underline{p}ercolation \underline{c}ommunity (\BCPC) is a widely studied approach for detecting cohesive structures in bipartite graphs \cite{lehmann2008biclique,jesus2009bipartite,liao2010visualizing,hecking2014analysis,hecking2015analysis,ziebarth2015resource,chen2023index}. A biclique is a complete bipartite subgraph, and a biclique percolation community is defined as a maximal set of bicliques that are connected through \textit{adjacency}. In the $ (\alpha,\beta) $-biclique percolation community, two bicliques are considered \textit{adjacent} if they share at least $ \alpha $ vertices on one side and $ \beta $ on the other (share an $ (\alpha,\beta) $-biclique). Biclique percolation community (\BCPC) has diverse applications across bipartite network scenarios. In social networks (users-entities as vertices, interaction as edges), \BCPC uncovers interest-based user groups to support interest or user recommendation \cite{hecking2015analysis}. In Wikipedia networks (articles-editors as vertices, editorial activities as edges), \BCPC identifies clusters, revealing topic-driven editor aggregation and coordinated content creation \cite{jesus2009bipartite}. For internet resource access (resources-visitors as vertices), \BCPC is used to analyze student-course resource relationships and their temporal evolution \cite{hecking2014analysis,ziebarth2015resource}. In enterprise heterogeneous graphs (hosts-users as vertices, interactions as edges), \BCPC is used to detect clusters to aid abnormal user behavior identification and malicious activity tracing for network security \cite{liao2010visualizing}. \BCPC allows overlaps between clusters, and this property is crucial because real-world users or entities are likely to belong to multiple communities. This property has also been extensively studied \cite{hecking2015analysis,jesus2009bipartite,hecking2014analysis,liao2010visualizing}.

Despite its widespread practical utility, the \BCPC method faces significant computational challenges due to its high time complexity. To the best of our knowledge, the existing \BCPC methods are all based on \cite{lehmann2008biclique}. Its framework was later implemented in \cite{wang2018bmtk} and used to build indexes for personalized search problems \cite{chen2023index}. The basic idea is based on \textit{\underline{m}aximal \underline{b}iclique \underline{a}djacency \underline{g}raph} (\MBAG). This method first enumerates all maximal bicliques and then uses them as vertices to construct a maximal biclique adjacency graph. In this graph, there is an edge between two vertices if and only if the overlapping part of the corresponding two maximal bicliques contains an $ (\alpha,\beta) $-biclique. The connected components in the adjacency graph correspond one-to-one with the $ (\alpha,\beta) $-\BCPC. However, real-world bipartite graphs may contain a large number of maximal bicliques (up to $ 2^{n/2} $ theoretically \cite{dai2023hereditary}), and the adjacency graph formed by these maximal bicliques could contain a significant number of edges (up to $ 10^{11} $, see Table~\ref{tab:datasets}). Traversing such a large-scale adjacency graph is highly time-consuming.

To overcome the limitations of existing methods, we first propose a partial-\BCPC based solution, which focuses on reducing the size of the \MBAG. The key idea is to leverage the prefixes of the maximal biclique enumeration tree during the enumeration process. Specifically, we introduce a novel concept of partial-\BCPC, which is computed on-the-fly while enumerating maximal bicliques. During the enumeration, if a prefix of the enumeration tree contains at least $ \alpha $ vertices from the vertex set $ U $ and $ \beta $ vertices from the vertex set $ V $, we can determine that all maximal bicliques under this branch belong to the same \BCPC. These bicliques collectively form an incomplete \BCPC, referred to as a partial-\BCPC. Maximal bicliques in the same partial-\BCPC can be regarded as a single vertex in the \MBAG, thus reducing the size of the \MBAG. Experiments show that our approach achieves a substantial reduction in the number of vertices in the \MBAG—by two orders of magnitude (see Figure~\ref{fig:pbcpc_num} (c) of Exp-3)—thereby enabling more efficient traversal and computation of \BCPCs.

While the partial-\BCPC solution operates on a reduced \MBAG, it still requires a full traversal of this structure, which can remain prohibitively large. To overcome this limitation, we propose a novel approach based on $(\alpha, \beta)$-biclique enumeration. The basic idea is to enumerate all $(\alpha,\beta)$-bicliques in the bipartite graph, and for each enumerated $(\alpha,\beta)$-biclique $B$, connect all maximal bicliques containing $B$ using union-find set. The reason this idea works is that two maximal bicliques are adjacent only if they share an $(\alpha,\beta)$-biclique. Since the number of $(\alpha, \beta)$-bicliques increases exponentially with respect to $ \alpha,\beta $ \cite{yang2023p,ye2023efficient}, we leverage partial-\BCPC once again to prune the $(\alpha, \beta)$-biclique enumeration tree. During the enumeration process, for each intermediate small biclique generated, we can maintain the set of maximal bicliques containing it. If this set is empty or all the maximal bicliques in the set already belong to the same partial-\BCPC, then the current enumeration branch can be safely pruned. Experiments show that using partial-\BCPC can reduce the number of nodes in the $(\alpha, \beta)$-biclique enumeration tree by more than five orders of magnitude (Exp-4). Consequently, it achieves a speedup of nearly three orders of magnitude compared to the state-of-the-art (SOTA) algorithm (Exp-1).

We summarize our contributions as follows.

\stitle{Novel partial-\BCPC based solution.} We propose a novel partial-\BCPC based solution, characterized by introducing a relaxed definition of \BCPC, namely partial-\BCPC. By leveraging partial-\BCPC, we can significantly reduce the size of the \MBAG used in existing methods, thereby directly accelerating the traversal process of the \MBAG. We show that leveraging partial-\BCPC can reduce the number of vertices in the \MBAG by up to two orders of magnitude (i.e., eliminating up to 99\% of the vertices in \MBAG, see Exp-3).

\stitle{Novel $ (\alpha,\beta) $-biclique based solution.} Unlike traditional approaches that rely on traversing the \MBAG, we propose a novel framework that directly enumerates $ (\alpha,\beta) $-bicliques and connects the maximal bicliques that share them utilizing union-find sets. To reduce the cost of enumerating $ (\alpha,\beta) $-bicliques, we again leverage partial-\BCPC, which reduces the number of nodes in the enumeration tree by up to five orders of magnitude (Exp-4).\comment{As a result, the overall algorithm achieves nearly three orders of magnitude improvement over the SOTA method (Exp-1).}

\stitle{Extensive experiments.} We conduct experiments on 10 real-world datasets, and the experimental results demonstrate the effectiveness and efficiency of our proposed algorithms. Specifically, we test the performance of our algorithms under a wide range of parameter settings. For example, on dataset \kw{Youtube}, when $ \alpha=2,\beta=2 $, the SOTA method (\MBAG) takes 13,376 seconds, while the partial-\BCPC based solution (\pbcpcp) takes 197 seconds, achieving an improvement of nearly two orders of magnitude. In contrast, the $ (\alpha,\beta) $-biclique based solution (\bicliquep) takes 34 seconds, with an improvement of nearly three orders of magnitude. Additionally, we conduct a case study of \BCPC in community detection. The results show that compared with other traditional community models (e.g., biclique, bitruss, bicore), \BCPC can find more practically meaningful communities, which demonstrates the effectiveness of \BCPC.

\section{Preliminaries}\label{sec:preliminaries}
Let $ G=(U,V,E) $ be an undirected and unweighted bipartite graph, where $ U,V $ are two disjoint sets of vertices and $ E $ is a set of edges in form of $ (u,v),u\in U,v\in V $. For a vertex $ u\in U $, $ Nei_G(u) $ is the set of neighbors of $ u $, i.e., $ Nei_G(u)=\{v|(u,v)\in E\} $. A bipartite graph $ G'=(U',V',E') $ is a subgraph of $ G $ if $ U'\subseteq U,V'\subseteq V,E'\subseteq E $.

\begin{definition}[Biclique]
	\label{def:biclique}
	Given a bipartite graph $ G=(U,V,E) $, a biclique is a pair of vertex sets $ B=(X,Y) $ where $ X\subseteq U,Y\subseteq V $ and $ \forall u\in X,v\in Y,(u,v)\in E $.
\end{definition}
For simplicity, we let $ B.X=X,B.Y=Y $ if $ B=(X,Y) $. For two bicliques $ B,B' $, $ B\subseteq B' $ indicates that $ B.X\subseteq B'.X,B.Y\subseteq B'.Y $. A maximal biclique $ B $ is a biclique that there is no other biclique $ B' $ satisfying $ B\subseteq B' $.

\begin{definition}[$ (\alpha,\beta) $-Biclique Adjacency]
	\label{def:bicliqueadj}
	Given two integer $ \alpha,\beta $ and two bicliques $ B_1,B_2 $, $ B_1 $ and $ B_2 $ are $ (\alpha,\beta) $-adjacent if $ |B_1.X\cap B_2.X|\geq \alpha,|B_1.Y\cap B_2.Y|\geq \beta $.
\end{definition}

\begin{definition}[$ (\alpha,\beta) $-Biclique Connectivity]
	\label{def:bicliqueconnect}
	Given two integer $ \alpha,\beta $ and two bicliques $ B_1,B_n $, $ B_1 $ and $ B_n $ are $ (\alpha,\beta) $-connected if there is a sequence of bicliques $ B_1,B_2,...,B_n $ that $ \forall i=1,...,n-1 $, $ B_i $ and $ B_{i+1} $ are $ (\alpha,\beta) $-adjacent.
\end{definition}

\begin{definition}[$ (\alpha,\beta) $-\underline{B}i\underline{c}lique \underline{P}ercolation \underline{C}ommunity\\($ (\alpha,\beta) $-\BCPC)]
	\label{def:bcpc}
	Given a bipartite graph $ G $ and two integer $ \alpha,\beta $, an $ (\alpha,\beta) $-biclique percolation community is a maximal set of maximal bicliques in $ G $ that any two maximal bicliques in the set are $ (\alpha,\beta) $-connected.
\end{definition}

\begin{figure}[t]
	\begin{center}
		\includegraphics[width=0.9\columnwidth]{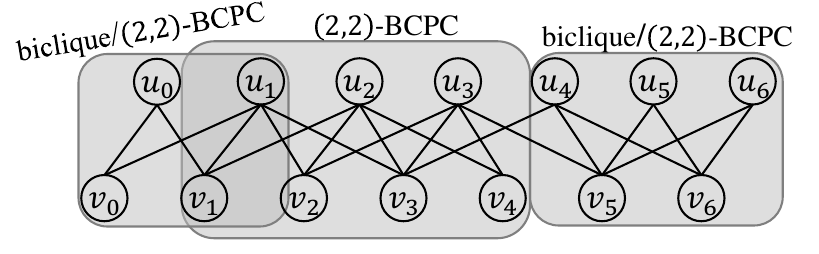}
	\end{center}
	\vspace*{-0.3cm}
	\caption{A bipartite graph $ G $ and $ (\alpha,\beta) $-\BCPC ($ \alpha=2,\beta=2 $)}
	\vspace*{-0.4cm}
	\label{fig:example_bigraph}
\end{figure}

\begin{example}\label{exm:bcpc_exm}
	Let $ \alpha=2,\beta=2 $, Figure~\ref{fig:example_bigraph} illustrates a bipartite graph \( G \) and some of the \( (2,2) \)-\BCPCs. These \BCPCs locates on the left, in the middle, and on the right, respectively. The \( (2,2) \)-\BCPC in the middle is composed of three maximal bicliques: \( B_1 = (\{u_1, u_2\}, \{v_1, v_2, v_3\}) \), \( B_2 = (\{u_1, u_2, u_3\}, \{v_2, v_3\}) \), and \( B_3 = (\{u_2, u_3\},\\\{v_2, v_3, v_4\}) \). Among these bicliques, \( B_1 \) and \( B_2 \) are adjacent because their intersection, \( (\{u_1, u_2\},\{v_2, v_3\}) \), satisfies the condition in Definition~\ref{def:bicliqueadj}. For the same reason, \( B_2 \) and \( B_3 \) are also adjacent. Therefore, \( B_1 \), \( B_2 \), and \( B_3 \) can together form a valid \( (2,2) \)-\BCPC. On the left and right sides of Figure~\ref{fig:example_bigraph} are two maximal bicliques. These maximal bicliques are not adjacent (Definition~\ref{def:bicliqueadj}) to any other maximal bicliques, and thus can be regarded as two special \BCPCs.
\end{example}

\stitle{Problem statement.} Given a bipartite graph $ G $ and two integer $ \alpha,\beta $, the goal of our work is to detect all $ (\alpha,\beta) $-\BCPCs (\BCPC detection).


\subsection{Maximal Biclique Enumeration}\label{subsec:mbe}

The existing state-of-the-art maximal biclique enumeration methods \cite{abidi2020pivot,chen2022efficient,dai2023hereditary} are all based on the set-enumeration tree structure, where vertices are iteratively traversed to build nodes of the enumeration tree, with each node representing a distinct biclique.\comment{These bicliques will be checked to determine whether they are maximal bicliques. In these methods, the pruning strategy with pivoting \cite{abidi2020pivot} and the vertex ordering strategy \cite{chen2022efficient} are both employed to enumerate maximal bicliques. \cite{dai2023hereditary} integrates the aforementioned strategies and proposes a unified framework that is applicable to the enumeration of all maximal hereditary subgraphs (including maximal bicliques) in bipartite graphs.} In this paper, we adopts the algorithmic framework in \cite{dai2023hereditary} to perform maximal biclique enumeration and to design the partial-\BCPC based method in Section~\ref{sec:new_mbag}. This is because the method in \cite{dai2023hereditary} achieves state-of-the-art performance in maximal biclique enumeration, and furthermore, the framework in \cite{dai2023hereditary} is adaptable to a wider range of cohesive subgraph models, making our partial-\BCPC based method in Section~\ref{sec:new_mbag} to also have the potential to be extended to other cohesive subgraph models in bipartite graphs.

\stitle{The basic framework \cite{dai2023hereditary}.} Algorithm~\ref{alg:mbe_frame} presents the basic framework for maximal biclique enumeration in \cite{dai2023hereditary}. This framework recursively explores candidate vertex sets to construct the set-enumeration tree, while leveraging exclusion sets to avoid non-maximal bicliques. Algorithm~\ref{alg:mbe_frame} begins by invoking the recursive procedure \kw{Enum} in Line~1. The parameters of \kw{Enum} contain the current biclique \( (R_U, R_V) \), candidate vertices \( (C_U, C_V) \) for expansion, and exclusion vertices \( (X_U, X_V) \) that have already been processed. Within \kw{Enum}, the algorithm first checks the termination condition: $ C_U\cup C_V=\emptyset $ (Line~3). If the termination condition is satisfied while $ X_U\cup X_V=\emptyset $ (Line~4), it implies that $ (R_U,R_V) $ is a maximal biclique. Subsequently, the algorithm iterates over each vertex \( u \) in the candidate set \( C_U \) (Line~6), calling the \kw{Branch} procedure to perform branch expansion (Line~7), and moves \( u \) from the candidate set \( C_U \) to the exclusion set \( X_U \) (Line~8). Symmetrically, the algorithm performs the same operation for each vertex \( v \) in \( C_V \) (Lines~9-11). In the \kw{Branch} procedure, the algorithm computes the new candidate set \( C'_V = C_V \cap Nei_G(u) \) and the new exclusion set \( X'_V = X_V \cap Nei_G(u) \), adds \( u \) to \( R_U \), and recursively calls the \kw{Enum} procedure to further explore maximal bicliques within this branch.

\begin{algorithm}[t]
	\scriptsize
	\caption{The basic maximal biclique enumeration framework in \cite{dai2023hereditary}}
	\label{alg:mbe_frame}
	\KwIn{Bipartite graph $ G=(U,V,E) $}
	\KwOut{All maximal bicliques in $ G $}
	\kw{Enum}$ (\emptyset,\emptyset,U,V,\emptyset,\emptyset) $;
	
	\vspace*{0.1cm}
	{\bf Procedure} {\kw{Enum}$ (R_U,R_V,C_U,C_V,X_U,X_V) $}
	
	\If{$ C_U\cup C_V=\emptyset $}{
		\lIf{$ X_U\cup X_V=\emptyset $}{Output $ (R_U,R_V) $ as a maximal biclique}
		\Return;
	}

	\ForEach{$ u\in C_U $}{
		\kw{Branch}$ (R_U,R_V,u,C_U-\{u\},C_V,X_U,X_V) $\;
		$ C_U\gets C_U-\{u\} $; $ X_U\gets X_U\cup \{u\} $;
	}

	\ForEach{$ v\in C_V $}{
		\kw{Branch}$ (R_V,R_U,v,C_V-\{v\},C_U,X_V,X_U) $\;
		$ C_V\gets C_V-\{v\} $; $ X_V\gets X_V\cup \{v\} $;
	}

	\vspace*{0.1cm}
	{\bf Procedure} {\kw{Branch}$ (R_U,R_V,u,C_U,C_V,X_U,X_V) $}
	
	$ C'_V\gets C_V\cap Nei_G(u) $; $ X'_V\gets X_V\cap Nei_G(u) $\;
	\kw{Enum}$ (R_U\cup \{u\},R_V,C_U,C'_V,X_U,X'_V) $;
	
\end{algorithm}

\stitle{Pivoting technique \cite{dai2023hereditary}.} A notable issue with the aforementioned framework is that it generates a significant amount of unnecessary computations that produce non-maximal bicliques. We first make the following observation: for any vertex \( u \in U \) in a bipartite graph \( G = (U, V) \), and for any maximal biclique \( B = (X, Y) \) in the bipartite graph, exactly one of the following two conditions must hold:
\begin{enumerate}
	\item \( B \) contains \( u \) (i.e., \( u \in X \)).
	\item \( B \) contains a vertex \( v \), where \( v \) is a non-neighbor of \( u \) in \( V \) (i.e., \( \exists v \in V - Nei_G(u), v \in Y \)).
\end{enumerate}

Based on the above observation, we can conclude that by selecting a pivot vertex \( u \) from \( X_U \), \( C_U \), \( X_V \), or \( C_V \), and enumerating only \( u \) (if \( u \notin X_U, X_V \)) and vertices in \( C_V \) (or \( C_U \), if \( u \in X_V, C_V \)) that are not neighbors of \( u \) in Lines 6-11 of Algorithm~\ref{alg:mbe_frame}, we can ensure the completeness of all maximal bicliques while reducing unnecessary enumeration branches that produce non-maximal bicliques. An effective method for selecting a pivot vertex is to choose the vertex in \( X_U \cup C_U \) (or \( X_V \cup C_V \)) that has the fewest non-neighbor vertices in \( C_V \) (or \( C_U \)) as the pivot vertex.

\stitle{Vertex ordering \cite{chen2022efficient}.} The efficiency of maximal biclique enumeration algorithms is often influenced by the order in which vertices are processed. \cite{dai2023hereditary} applied the 2-hop degree ordering introduced in \cite{chen2022efficient} to their framework. This vertex ordering is only applied at the first level of the recursive search, while the pivoting pruning technique mentioned earlier is utilized in subsequent levels of the recursive search.

\begin{figure}[t]
	\begin{center}
		\includegraphics[width=0.95\columnwidth]{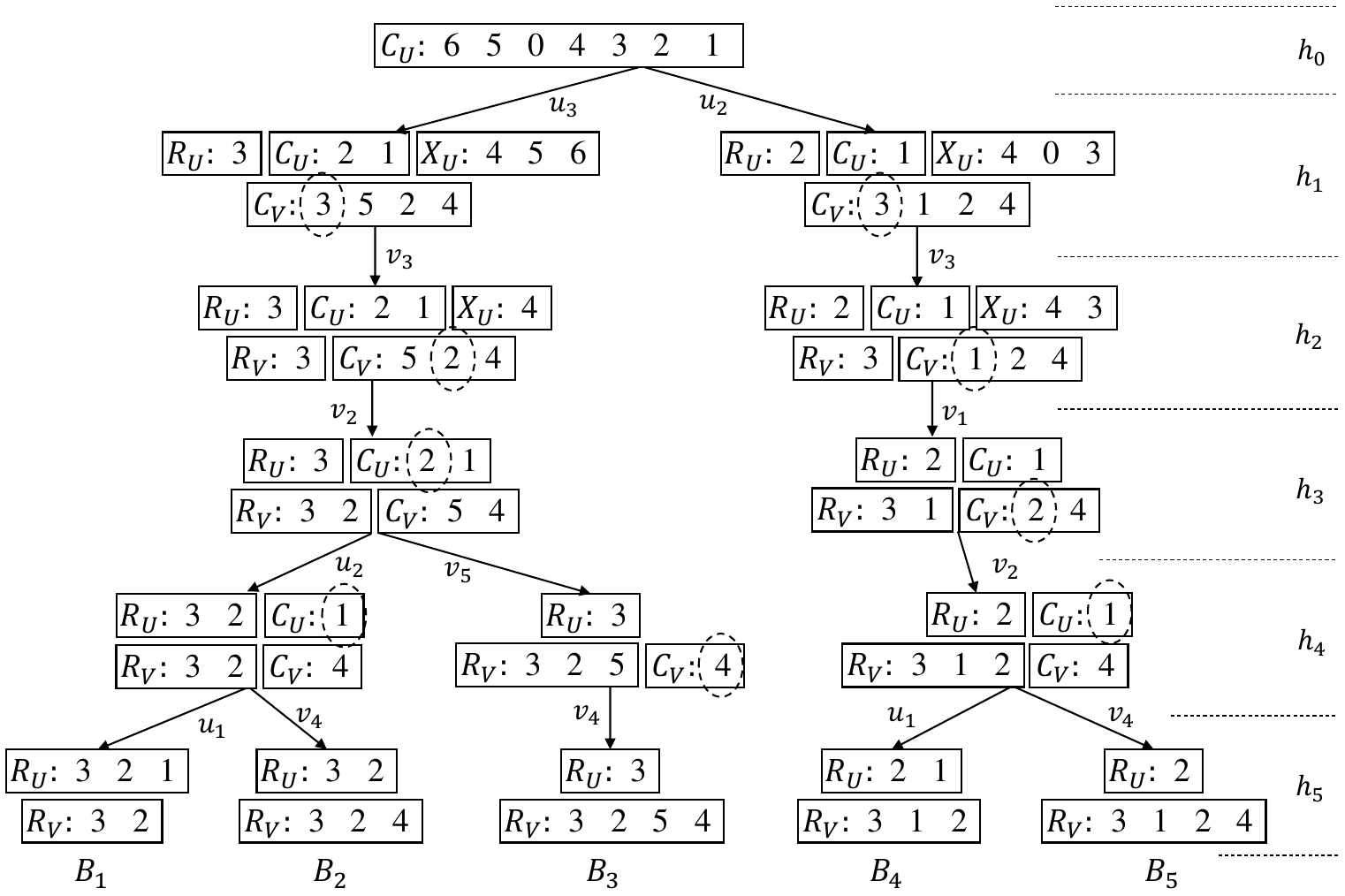}
	\end{center}
	\vspace*{-0.3cm}
	\caption{The partial enumeration tree of the framework in \cite{dai2023hereditary} on bipartite graph of Figure~\ref{fig:example_bigraph}. Pivot vertices are enclosed in dashed circles. \( nd(B_i, h_j) \) represents the tree node at level \( h_j \) on the \( B_i \) branch of the enumeration tree}
	\vspace*{-0.4cm}
	\label{fig:example_mbetree}
\end{figure}

\begin{example}\label{exm:mbexm}
	Figure~\ref{fig:example_mbetree} illustrates a partial enumeration tree for maximal biclique enumeration \cite{dai2023hereditary} on bipartite graph $ G=(U,V,E) $ in Figure~\ref{fig:example_bigraph}. The root node, located at level \( h_0 \), initializes its candidate set \( C_U \) with vertex set \( U \) of the bipartite graph, ordered by 2-hop degree ordering. The root node only contains \( C_U \) and no \( C_V \). The enumeration process in level $ h_0 $ follows the ordered sequence, processing \( u_6 \), \( u_5 \), \( u_0 \), and \( u_4 \) first, followed by \( u_3 \) and \( u_2 \), while only branches processing \( u_3 \) and \( u_2 \) are shown. For tree node $ nd(B_1,h_1) $, \( u_3 \) is added to \( R_U \), and \( C_V \) is updated to include $ Nei_G(u_3) $ in \( V \). The exclusion set \( X_U \) in $ nd(B_1,h_1) $ contains \( u_4 \), \( u_5 \), \( u_6 \), as these vertices were enumerated in previous branches and added to \( X_U \) at level \( h_0 \) (initially, \( X_U \) at level \( h_0 \) is empty and is not shown). 
	
	At $ nd(B_1,h_1) $, the framework selects \( v_3 \) as the pivot vertex. Consequently, vertices in \( C_V \) other than \( v_3 \) and vertices in $ Nei_G(v_3)\cap C_U $ (\( u_2 \), \( u_1 \)) are skipped. Thus, only the branch enumerating \( v_3 \) is expanded at level \( h_2 \). At level \( h_2 \), \( v_3 \) is added to \( R_V \), and removed from \( C_V \). The sets \( C_U \) and \( X_U \) at level \( h_2 \) are updated similarly to further refine the search space and exclusion set. When both candidate sets \( C_U \) and \( C_V \) are empty (i.e., at the leaf nodes of the enumeration tree), and if the exclusion sets \( X_U \) and \( X_V \) are also empty, then \( (R_U, R_V) \) is a maximal biclique.
\end{example}

\subsection{$ (\alpha,\beta) $-Biclique Enumeration}\label{subsec:abbcl}
The existing state-of-the-art $ (\alpha,\beta) $-biclique enumeration method is proposed in \cite{yang2023p}. Given a bipartite graph $ G=(U,V,E) $, the core idea is to enumerate combinations of vertices in the $U$ set while maintaining their common neighbors in the $V$ set. It also constructs an ordinary directed graph, called the \textit{2-hop graph}, on the vertices of the $U$ set to accelerate the enumeration process.

Algorithm~\ref{alg:ablist_frame} presents the basic framework. It first constructs the \textit{2-hop graph} $ \vec{H} $ with $ U $ as its vertex set (Line~1), where an edge $ (u,v) $ exists if $ u<v $ and $ u,v $ share at least one common neighbor in set $ V $ of the bipartite graph. In Line~4, if $R_U$ already contains $\alpha$ vertices, it is sufficient to enumerate all combinations of $\beta$ vertices from $C_V$ to obtain the $(\alpha,\beta)$-bicliques. In Line~7, the algorithm enumerates vertices $u$ in $\vec{H}$, then in Line~8 updates the candidate set $C_V'$ in $V$, and proceeds with the next-level enumeration in Line~11.

\begin{algorithm}[t]
	\scriptsize
	\caption{The $ (\alpha,\beta) $-biclique enumeration framework in \cite{yang2023p}}
	\label{alg:ablist_frame}
	\KwIn{Bipartite graph $ G=(U,V,E) $; Two integer $ \alpha,\beta $}
	\KwOut{All $ (\alpha,\beta) $-bicliques in $ G $}
	
	Construct 2-hop graph $ \vec{H}=(U,\vec{E_h}) $ on vertex set $ U $ \cite{yang2023p}; \tcc{$ \vec{H}=(U,\vec{E_h}) $: a direct graph, $ (u,v)\in \vec{E_h} $ indicates that $ u<v $, $ u,v $ have common neighbors in $ V $}
	
	\kw{BicliqueListing}$ (\vec{H},\emptyset,V) $\;
	
	\vspace*{0.1cm}
	{\bf Procedure} {\kw{BicliqueListing}$ (\vec{H},R_U,C_V) $}
	
	\If{$ |R_U|=\alpha $}{
		\lForEach{$ R_V\subseteq C_V,|R_V|=\beta $}{
			Output $ (R_U,R_V) $ as an $ (\alpha,\beta) $-biclique
		}
		\Return;
	}
	
	\ForEach{$ u\in \vec{H} $}{
		$ C_V'\gets C_V\cap Nei_G(u) $\;
		\lIf{$ |C_V'|<\beta $ or $ Nei_{\vec{H}}(u)<\alpha-|R_U|-1 $}{\textbf{continue}}
		
		Let $ \vec{H}' $ be induced subgraph of $ \vec{H} $ on $ Nei_{\vec{H}}(u) $\;
		\kw{BicliqueListing}$ (\vec{H}',R_U\cup \{u\},C_V') $;
	}
	
\end{algorithm}

\subsection{Union-Find Set \cite{cormen2022introduction}}\label{subsec:unionfind}
The Union-Find Set, also referred to as the Disjoint-Set data structure, serves as an efficient method for managing a collection of disjoint sets. Each set is represented as a tree, where the nodes represent the elements of the set. Within this tree, every node points to its parent, with the root of the tree acting as the representative of the set. This tree structure allows for fast \textbf{union} and \textbf{find} operations, making the Union-Find Set well-suited for tasks such as graph connectivity problems. \textbf{Find($i$)} is used to get the representative of the unique set containing $i$, \textbf{union($i,j$)} is used to merge the two sets containing \(i\) and \(j\) into their union. Specifically, within the corresponding tree structure, the representative node of one set is assigned as the parent of the representative node of another set.

In all practical scenarios, the \textbf{union} and \textbf{find} operations of the union-find set achieve constant amortized time complexity. The space complexity of this structure is \(O(n)\), with \(n\) denoting the total number of elements it manages \cite{cormen2022introduction}.

\def\MBE{\kw{MBE}}

\section{Existing Solution}\label{sec:exist_solution}
The existing \BCPC detection method was proposed and implemented in \cite{lehmann2008biclique,wang2018bmtk,chen2023index}. To the best of our knowledge, no other work has primarily focused on the algorithmic implementation and improvement of \BCPC detection. The basic idea is based on \textit{\underline{m}aximal \underline{b}iclique \underline{a}djacency \underline{g}raph} (\MBAG). This method first enumerates all maximal bicliques and then uses them as vertices to construct a maximal biclique adjacency graph. In this graph, there is an edge between two vertices if and only if the overlapping part of the corresponding two maximal bicliques contains an $ (\alpha,\beta) $-biclique (satisfying Definition~\ref{def:bicliqueadj}). The connected components in the adjacency graph correspond one-to-one with the $ (\alpha,\beta) $-\BCPC. In this method, the $ (\alpha,\beta) $-biclique connectivity of maximal bicliques in $ (\alpha,\beta) $-\BCPC are maintained by the classic union-find set \cite{cormen2022introduction}.




\begin{algorithm}[t]
	\scriptsize
	\caption{Existing \MBAG-based solution}
	\label{alg:bcpc_baseline}
	\KwIn{Bipartite graph $ G=(U,V,E) $; Two integer $ \alpha,\beta $}
	\KwOut{All \BCPC in $ G $}
	$ \Br\gets $ set of all maximal bicliques\;
	$ \Br\gets \{B|B\in \Br,|B.X|\geq \alpha,|B.Y|\geq \beta\} $\;
	$ Nei(B)\gets $ set of maximal bicliques that share vertices with $ B $\;
	$ UF\gets $ an initial union-find set, i.e., $ UF.find(B)=B $ for any input $ B $\;
	
	\For{$ B\in \Br $}{
		\For{$ B'\in Nei(B) $}{
			\lIf{$ |B.X\cap B'.X|\geq \alpha $ and $ |B.Y\cap B'.Y|\geq \beta $}{$ UF.union(B,B') $}
		}
	}

	\Return $ UF $;
	
\end{algorithm}

Algorithm~\ref{alg:bcpc_baseline} is the existing \MBAG-based algorithm identifying all $(\alpha,\beta)$-\BCPCs in a bipartite graph \( G = (U, V, E) \). It starts by enumerating all maximal bicliques in \( G \) and filters them based on size constraints (Lines~1-2). The reason is that for a maximal biclique $B$, if $|B.X|<\alpha$ or $|B.Y|<\beta$, then $B$ can not form a \BCPC with other maximal bicliques. A union-find set \( UF \) is initialized in Line~4. The algorithm then iterates over each maximal biclique \( B \) and its neighbors \( B'\in Nei(B) \), checking if \( |B.X\cap B'.X|\geq \alpha,|B.Y\cap B'.Y|\geq \beta \) (Lines~5-7). If the conditions are satisfied, \( B \) and \( B' \) are merged into the same \BCPC by $UF.union(\ast)$ (Line~7). After processing all maximal bicliques, the union-find set \( UF \) represents the final \BCPCs in the bipartite graph.

The major drawback of this approach is that the size of the \MBAG can be extremely large (up to $ 10^{11} $, see Table~\ref{tab:datasets}). Consequently, traversing such a massive \MBAG to derive all \BCPCs is clearly impractical.

\section{Novel Partial-\BCPC Based Solution}\label{sec:new_mbag}

As we discussed in Section~\ref{sec:exist_solution}, current approach relying on \MBAG (Algorithm~\ref{alg:bcpc_baseline}) faces significant challenges due to the vast scale of \MBAG. This immense size leads to inefficient traversal operations. Consequently, it becomes evident that optimizing the performance of \MBAG-based method hinges on effectively reducing the size of the \MBAG.

\stitle{Key idea.} We reduce the size of the \MBAG by merging vertices within it. The key is to leverage the prefixes of the maximal biclique enumeration tree during the enumeration process. During the enumeration, if a prefix of the enumeration tree contains at least $ \alpha $ vertices in $ U $ and $ \beta $ vertices in $ V $, we can determine that all maximal bicliques under this branch belong to the same \BCPC. Connections between these maximal bicliques are unnecessary to be traversed. These bicliques collectively form an incomplete \BCPC, referred to as a partial-\BCPC.\comment{By identifying and utilizing such partial-\BCPCs in the enumeration process, we can significantly reduce the size of the \MBAG, as these partial structures eliminate the need to explicitly represent and traverse redundant connections of maximal bicliques in the same partial-\BCPC. This is particularly advantageous for large-scale bipartite graphs, where the size of the \MBAG is a critical bottleneck.}

\begin{example}\label{exm:keyidea_example}
	Let $\alpha = \beta = 1$. Consider Figure~\ref{fig:example_mbetree}, where the subtree rooted at node $nd(B_1, h_2)$ contains three maximal biclique branches: $B_1$, $B_2$, and $B_3$. All three maximal bicliques include both $nd(B_1, h_2).R_U$ and $nd(B_1, h_2).R_V$. Therefore, we can regard $B_1$, $B_2$, and $B_3$ as forming an incomplete \BCPC, i.e., a partial-\BCPC.
\end{example}

Next, we will formally define partial-\BCPC, the maximal biclique enumeration tree and the special prefixes used to compute partial-\BCPCs.

\begin{definition}[partial-\BCPC]
	\label{def:par_bcpc}
	Given a bipartite graph $ G $ and two integer $ \alpha,\beta $, a partial-\BCPC $ \PPr $ is a set of maximal bicliques satisfying $ \PPr\subseteq \Br $, where $ \Br $ is an $ (\alpha,\beta) $-\BCPC in $ G $.
\end{definition}

The distinction from Definition~\ref{def:bcpc} lies in the fact that Definition~\ref{def:par_bcpc} does not require the set of maximal bicliques in the partial-\BCPC to be maximal. This reflects the characteristic of partial-\BCPC as an incomplete \BCPC. In practice, partial-\BCPCs are also maintained by union-find set.

\begin{definition}[\underline{M}aximal \underline{B}iclique \underline{E}numeration Tree (\MBE Tree)]
	\label{def:mbetree}
	Given a bipartite graph \( G = (U, V, E) \). The \MBE tree \( T = (N, \vec{E}) \) represents the search space of the maximal biclique enumeration algorithm on \( G \). Here, \( N \) denotes the set of nodes in the tree, where each node \( nd \in N \) contains a six-tuple \( (R_U, R_V, C_U, C_V, X_U, X_V) \), as defined in the basic framework of \MBE in Section~\ref{subsec:mbe}. The set \( \vec{E} \) consists of directed edges of the form \( (nd_i, nd_j, u) \), where \( u \in U \) or $ u\in V $. An edge \( (nd_i, nd_j, u) \in \vec{E} \) indicates that node \( nd_j \) is a child of node \( nd_i \), generated by selecting vertex \( u \) from \( nd_i.C_U \) or \( nd_i.C_V \). A node $ nd\in N $ is a leaf node if $ nd.C_U\cup nd.C_V=\emptyset $. For a leaf node $ nd\in N $, if $ nd.X_U\cup nd.X_V=\emptyset $, then $ (nd.R_U,nd.R_V) $ is a maximal biclique.
\end{definition}

For simplicity, we use $nd.RCX_U$ and $nd.RCX_V$ to denote $nd.R_U \cup nd.C_U \cup nd.X_U$ and $nd.R_V \cup nd.C_V \cup nd.X_V$, respectively.

\begin{definition}[$ (\alpha,\beta) $-Prefix and $ (\alpha,\beta) $-Node]
	\label{def:abprefix}
	Given a bipartite graph \( G = (U, V, E) \) and its \MBE tree \( T = (N, \vec{E}) \), let $ nd_1\in N $ be the root node of $ T $, $ PF(nd_l)=(nd_1,nd_2,...,nd_l) $ is the path from $ nd_1 $ to $ nd_l $, i.e., $ \forall i=1,2,...,l-1, (nd_i,nd_{i+1},\ast)\in \vec{E} $. $ PF(nd_l) $ is an $ (\alpha,\beta) $-prefix if and only if $ |nd_l.R_U|=\alpha,|nd_{l-1}.R_U|<\alpha,|nd_l.R_V|\geq \beta $ or $ |nd_l.R_V|=\beta,|nd_{l-1}.R_V|<\beta,|nd_l.R_U|\geq \alpha $. In this case, $ nd_l $ is an $ (\alpha,\beta) $-node.
\end{definition}

In short, an \( (\alpha, \beta) \)-node \( nd \) is the first node that satisfies \( |nd.R_U| \geq \alpha \) and \( |nd.R_V| \geq \beta \) in all branches where \( nd \) resides. For all its ancestor nodes \( nd_i \), it is not possible to simultaneously satisfy \( |nd_i.R_U| \geq \alpha \) and \( |nd_i.R_V| \geq \beta \).

\begin{theorem}\label{thm:abnode_nonancestor}
	For any two \( (\alpha, \beta) \)-nodes $ nd_1,nd_2 $, $ nd_1 $ can not be an ancestor or a descendant of $ nd_2 $.
\end{theorem}
\ifdefined\showproof
\begin{proof}
	Based on Definition~\ref{def:abprefix}, $ nd_1 $ is the first node that satisfies \( |nd_1.R_U| \geq \alpha \) and \( |nd_1.R_V| \geq \beta \) in all branches where \( nd_1 \) resides. The same applies to $ nd_2 $. Thus, $ nd_1,nd_2 $ are in two different branches, and $ nd_1 $ can not be an ancestor or a descendant of $ nd_2 $.
\end{proof}
\else
\fi

\subsection{The Basic Framework}

This section presents the basic framework for our partial-\BCPC based solution. The framework consists of two fundamental steps: (1) Computing partial-\BCPCs using the \MBE tree; (2) Traversing the reduced \MBAG based on the partial-\BCPCs to derive the final \BCPCs. When computing partial-\BCPCs, the framework connects the maximal bicliques within the subtree rooted at specific nodes in the \MBE tree using a union-find set. Specifically, we focus on $(\alpha, \beta)$-nodes in \MBE tree. Ancestors of an $(\alpha, \beta)$-node are ignored as they fail to satisfy $|R_U| \geq \alpha$ and $|R_V| \geq \beta$. Descendants of an $(\alpha, \beta)$-node, while satisfying $|R_U| \geq \alpha$ and $|R_V| \geq \beta$, are redundant since any maximal biclique branches sharing the prefix corresponding to the descendants must also share the \( (\alpha, \beta) \)-prefix corresponding to the \( (\alpha, \beta) \)-node. Thus, focusing solely on $(\alpha, \beta)$-nodes enables efficient computation of partial-\BCPCs.

\begin{algorithm}[t]
	\scriptsize
	\caption{Partial-\BCPC based solution}
	\label{alg:bcpc_par_bcpc_basic}
	\KwIn{Bipartite graph $ G=(U,V,E) $; Two integer $ \alpha,\beta $}
	\KwOut{All \BCPC in $ G $}
	Let $ T=(N,\vec{E}) $ be the \MBE tree of $ G $, $ \Br $ be the set of maximal bicliques in $ G $\;
	$ Nei(B)\gets $ set of maximal bicliques that share vertices with $ B $\;
	Let $ UF $ be an initial union-find set\;
	Let $\PPr \gets \emptyset$\;
	\tcc{compute partial-\BCPCs}
	\For{$(\alpha, \beta)$-node $nd$ in $T$}{
		\kw{ProcessNode}$ (nd) $\;
		\lFor{$ B\in \PPr $}{$ UF.union(B,\PPr[0]) $}
		$ \PPr\gets \emptyset $;
	}
	\tcc{traverse \MBAG with partial-\BCPCs}
	\For{$ B\in \Br $}{
		\For{$ B'\in Nei(B) $ and $ B,B' $ are in different partial-\BCPCs}{
			\lIf{$ |B.X\cap B'.X|\geq \alpha,|B.Y\cap B'.Y|\geq \beta $}{
				$ UF.union(B,B') $
			}
		}
	}
	
	\Return $ UF $;
	
	\vspace*{0.1cm}
	\tcc{recursively traverse the \MBE tree to gather all maximal bicliques under $ node $}
	{\bf Procedure} {\kw{ProcessNode}$ (node) $}
	
	\For{$ (node,node',u)\in \vec{E} $}{
		\If{$ node'.C_U\cup node'.C_V=\emptyset $}{
			\If{$ node'.X_U\cup node'.X_V=\emptyset $}{
				$ \PPr\gets \PPr\cup \{(node'.R_U,node'.R_V)\} $; \tcc{$ (node'.R_U,node'.R_V) $ is a maximal biclique}
			}
		}
		\lElse{
			\kw{ProcessNode}$ (node') $
		}
		
	}
	
\end{algorithm}

\stitle{Implementation.} Algorithm~\ref{alg:bcpc_par_bcpc_basic} presents the basic framework. It begins by maximal biclique enumeration and extracting \MBE tree \( T = (N, \vec{E}) \) (Line~1). It then initializes the neighbor list of maximal bicliques and a union-find set \( UF \) to manage the partial-\BCPCs and \BCPCs (Lines~2-3). In Lines~5-8, the algorithm iterates over each \( (\alpha, \beta) \)-node \( nd \) in \( T \) (Line~5), invoking the \kw{ProcessNode} function to gather all maximal bicliques under subtree rooted at $ nd $ into \( \PPr \) (Line~6). Subsequently, it merges all maximal bicliques in \( \PPr \) into the same partial-\BCPC (Line~7). In Lines~9-11, the algorithm traverses a smaller \MBAG based on the partial-\BCPCs. Specifically, for each maximal biclique \( B \), the algorithm only traverses maximal bicliques \( B' \) that reside in different partial-\BCPCs from \( B \) (Line~10). The algorithm finally returns \( UF \), which represents all detected \BCPCs.

\stitle{The storage of the \MBE tree.} In practice, we do not store all branches of the \MBE tree. We first introduce the two types of nodes in the \MBE tree.
\begin{definition}[Real and Virtual Node]
	\label{def:realvirtualnode_cp}
	Given a graph $G$ and its \MBE tree $ T=(N,\vec{E}) $, for any node $nd\in N$, if $nd$ is a leaf node and satisfies $nd.X_U = nd.X_V = \emptyset$, then $nd$ is a real node—in this case, $(nd.R_U, nd.R_V)$ is a maximal biclique (see Definition~\ref{def:mbetree}). Otherwise, if $nd$ has at least one descendant that is a real node, it is also a real node. If neither of these conditions is satisfied, then $nd$ is a virtual node.
\end{definition}

The \kw{ProcessNode} procedure in Algorithm~\ref{alg:bcpc_par_bcpc_basic} only needs to identify the branches that contain maximal bicliques (Lines~15-17), i.e., branches composed of real nodes. Thus, in practice, we only store real nodes in the \MBE tree.

\stitle{Analysis.} We first analyze the correctness of Algorithm~\ref{alg:bcpc_par_bcpc_basic}, then examine the relationship between the number of partial-\BCPCs obtained by Algorithm~\ref{alg:bcpc_par_bcpc_basic} and the numbers of maximal bicliques as well as \BCPCs, and finally analyze the complexity of Algorithm~\ref{alg:bcpc_par_bcpc_basic}.

\begin{theorem}\label{thm:pbcpc_basic_correct}
	Algorithm~\ref{alg:bcpc_par_bcpc_basic} correctly computes all \BCPCs.
\end{theorem}

\ifdefined\showproof
\begin{proof}
	We first prove that the partial-\BCPCs are all correct. The procedure \kw{ProcessNode} collects all maximal bicliques within the subtree rooted at the given $ node $ into \PPr. Since the parameter is always an $(\alpha, \beta)$-node, the maximal bicliques collected in each execution of the procedure must share an $(\alpha, \beta)$-biclique. Therefore, these bicliques correctly form a partial-\BCPC. Then we prove that the final \BCPCs are correct. In Lines~9-11, any two maximal bicliques that belong to different partial-\BCPCs but satisfy the adjacency condition are connected via the union-find set $ UF $. As a result, each disjoint set maintained by $ UF $ corresponds to a valid \BCPC.
\end{proof}
\else

\fi

\begin{theorem}\label{thm:pbcpc_basic_num}
	Given a bipartite graph $ G $, two integer $ \alpha,\beta $ and \MBE tree $ T=(N,\vec{E}) $, let $ n_{biclique} $ be the number of maximal bicliques, $ ND\gets \{nd|nd\in N$, nd is an $ (\alpha,\beta) $-node, nd is a real node$\} $, $ pbcpc $ be the number of partial-\BCPCs computed by Algorithm~\ref{alg:bcpc_par_bcpc_basic}, $ bcpc $ be the number of \BCPCs, the following inequality holds: $ n_{biclique}\geq |ND|=pbcpc\geq bcpc $.
\end{theorem}
\ifdefined\showproof
\begin{proof}
	Based on Theorem~\ref{thm:abnode_nonancestor}, any two $ (\alpha,\beta) $-nodes in $ ND $ can not be in the same branch, and nodes in $ ND $ are all real nodes, thus, each node in $ ND $ has at least one maximal biclique in its subtree. As a result, $ n_{biclique}\geq |ND| $. Since Algorithm~\ref{alg:bcpc_par_bcpc_basic} connects all maximal bicliques in the subtree of each $(\alpha,\beta)$-node in $ ND $ into a partial-\BCPC, and the maximal bicliques from different subtrees of $(\alpha,\beta)$-nodes are distinct, thus, each $(\alpha,\beta)$-node in $ ND $ corresponds to one partial-\BCPC ($ |ND|=pbcpc $). Partial-\BCPCs are incomplete \BCPCs according to Definition~\ref{def:par_bcpc}, thus, $ pbcpc\geq bcpc $.
\end{proof}
\else
\fi

\begin{theorem}\label{thm:pbcpc_time_space}
	The time and space complexity of Algorithm~\ref{alg:bcpc_par_bcpc_basic} are $ O((1-\frac{1}{pbcpc})n_{biclique}^2n) $ and $ O(n_{biclique}n) $ respectively, where $ n=|U|+|V| $, $ pbcpc $ is the number of partial-\BCPCs, $ n_{biclique} $ is the number of maximal bicliques.
\end{theorem}
\ifdefined\showproof
\begin{proof}
	Let $ m=|E| $, the time complexity of maximal biclique enumeration is $ O(2^{n/2}m) $ \cite{dai2023hereditary}. Compared to maximal biclique enumeration, Algorithm~\ref{alg:bcpc_par_bcpc_basic} introduces additional operations: traversing $(\alpha,\beta)$-nodes and collecting maximal biclique branches from the subtrees rooted at these nodes. Since $(\alpha,\beta)$-nodes do not reside in the same branch (Theorem~\ref{thm:abnode_nonancestor}), the time complexity of these additional operations does not exceed $O(2^{n/2}m)$. In Lines~9-11, in the worst case, Algorithm~\ref{alg:bcpc_par_bcpc_basic} only skips the inner mutual visits of maximal bicliques in each initial partial-\BCPC. Let $ x=n_{biclique},y=pbcpc $ and $ p_1,p_2,...,p_y $ be the number of maximal bicliques in each partial-\BCPC, the time complexity of Lines~9-11 is $ O(x^2n-(p_1^2+p_2^2+...+p_y^2)n)=O(x^2n-(\frac{x}{y})^2yn) $ (based on Cauchy-Schwarz inequality), which can be written as $ O((1-\frac{1}{pbcpc})n_{biclique}^2n) $. As a result, the time complexity of Algorithm~\ref{alg:bcpc_par_bcpc_basic} is $ O(2^{n/2}m+(1-\frac{1}{pbcpc})n_{biclique}^2n) $. Since $ O(n_{biclique})=O(2^{n/2}) $ \cite{prisner2000bicliques,dai2023hereditary}, the time complexity can be rewritten as $ O((1-\frac{1}{pbcpc})n_{biclique}^2n) $.
	
	As for the space complexity, the main space consumption in Algorithm~\ref{alg:bcpc_par_bcpc_basic} comes from the \MBE tree and maximal bicliques. Since only the real nodes in the \MBE tree are stored, the number of branches in the \MBE tree is equal to the number of maximal bicliques, thus, the space complexity of Algorithm~\ref{alg:bcpc_par_bcpc_basic} is $ O(n_{biclique}n) $.

\end{proof}
\else
\fi

\stitle{Discussion.} The inequality $n_{biclique} \geq pbcpc$ in Theorem~\ref{thm:pbcpc_basic_num} forms the basis for reducing the \MBAG using partial-\BCPC. This is because maximal bicliques within the same partial-\BCPC are treated as a single vertex in the \MBAG, which leads to an \MBAG with formally fewer vertices. This can also be seen from the time complexity in Theorem~\ref{thm:pbcpc_time_space}: if $ pbcpc $ is replaced with $ n_{biclique} $, the result is exactly the complexity of the existing method in Section~\ref{sec:exist_solution} for traversing the complete \MBAG ($ O(n_{biclique}^2n) $). Experiments (Figure~\ref{fig:pbcpc_num} (c)) show that $ pbcpc $ can be an order of magnitude smaller than $ n_{biclique} $.

\subsection{Optimizations}
\stitle{Beyond subtree rooted at $(\alpha, \beta)$-node.} The basic framework in the previous section directly merges the maximal bicliques in the subtree rooted at an $(\alpha, \beta)$-node into a partial-\BCPC. However, this approach is insufficient because, for an $(\alpha, \beta)$-node $nd$, not all maximal bicliques sharing $nd.R_U$ and $nd.R_V$ are located within the subtree rooted at $nd$. For example, in Figure~\ref{fig:example_mbetree}, if $\alpha = \beta = 1$, $nd(B_4, h_2)$ is an $(\alpha, \beta)$-node, but $B_1$ and $B_2$ are not in the subtree rooted at $nd(B_4, h_2)$, even though both $B_1$ and $B_2$ share $nd(B_4, h_2).R_U$ and $nd(B_4, h_2).R_V$. This phenomenon occurs because some vertices in $B_1$ and $B_2$ (e.g., $ u_3 $) appear early in the candidate set $C_U$ of the root node, causing $B_1$ and $B_2$ to be enumerated much earlier.

To merge all relevant maximal bicliques within and beyond the subtree of an $(\alpha, \beta)$-node, it is essential to fully utilize all the information maintained in the $(\alpha, \beta)$-node, i.e., $R_U$, $R_V$, $C_U$, $C_V$, $X_U$, and $X_V$, to systematically search on the \MBE tree. Specifically, for each $(\alpha, \beta)$-node $nd$, we use $nd.RCX_U$ and $nd.RCX_V$ to search, within the \MBE tree, all branches of maximal bicliques that contain $nd.R_U$ and $nd.R_V$, and connect them into the same partial-\BCPC.

\begin{figure}[t]
	\begin{center}
		\begin{tabular}[t]{c}
			\hspace*{-0cm}
			\subfigure[{\scriptsize Exploring both branches within and beyond subtree rooted at \(nd\)}]{
				\includegraphics[width=0.99\columnwidth]{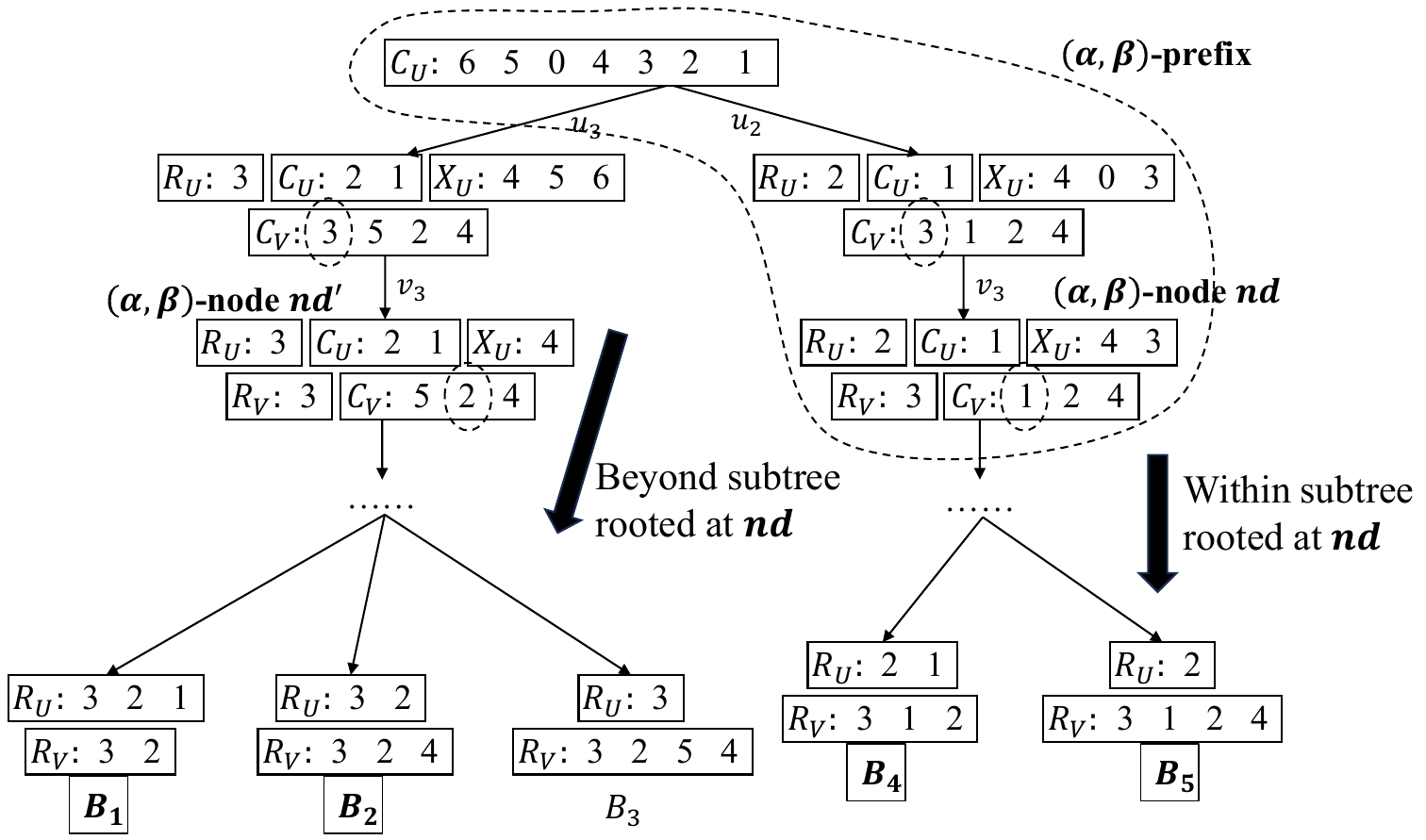}
			}
			\\
			\subfigure[{\scriptsize Using stop-labels to terminate exploration early}]{
				\includegraphics[width=0.95\columnwidth]{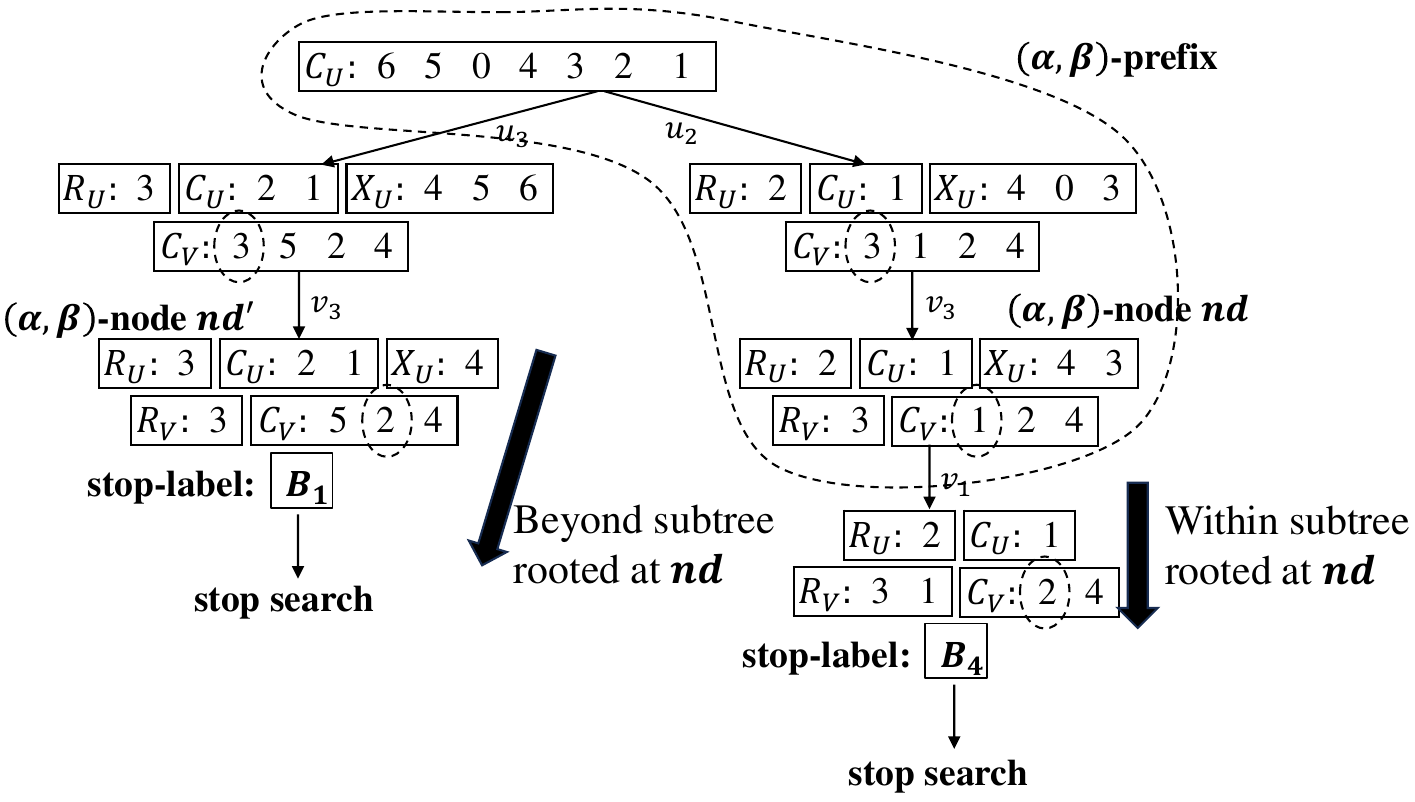}
			}
		\end{tabular}
	\end{center}
	\vspace*{-0.3cm}
	\caption{Examples of optimization}
	\vspace*{-0.4cm}
	\label{fig:example_beyond_stoplabel}
\end{figure}

\begin{example}\label{exm:beyond_subtree}
	Let $ \alpha=1,\beta=1 $, Figure~\ref{fig:example_beyond_stoplabel} (a) illustrates that when processing \((\alpha, \beta)\)-node \(nd\), both the branches within and beyond subtree rooted at \(nd\) are taken into account. It can be observed that the maximal biclique branches sharing \((nd.R_U, nd.R_V)\) include not only \(B_4\) and \(B_5\) under \(nd\) but also the maximal biclique branches \(B_1\) and \(B_2\) beyond the subtree rooted at $ nd $.
\end{example}

\stitle{Stop-Label.} The above method may result in fully traversing all the relevant maximal biclique branches when processing each $(\alpha, \beta)$-node. To reduce the overhead of traversing all maximal biclique branches that share $nd.R_U$ and $nd.R_V$ ($ nd $ is an \( (\alpha, \beta) \)-node) without affecting correctness, we assign a stop-label to certain real nodes in the \MBE tree. When such a real node is visited, the stop-label allows immediate retrieval of all maximal biclique information within its subtree, thus eliminating the need to further traverse its descendants. Given $ \alpha,\beta $, the following presents the details of the stop-label:
\begin{enumerate}
	\item Stop-labels are assigned only to real nodes in the \MBE tree.
	\item For a descendant node $nd$ of an $(\alpha, \beta)$-node, the prerequisite for setting stop-label to $nd$ is to connect all maximal bicliques within the subtree rooted at $nd$ into the same partial-\BCPC. The stop-label of $ nd $ is set to one of the maximal bicliques in that partial-\BCPC.\comment{In detail, if $nd$ is a leaf node, its stop-label is simply the maximal biclique represented by $nd$. If $nd$ is not a leaf node, the stop-labels of all its child nodes (also real nodes) are first connected into the same partial-\BCPC using union-find set. Then, the stop-label of $nd$ is set to any one of the maximal bicliques within that partial-\BCPC.}
	\item For an $(\alpha, \beta)$-node $nd$, the prerequisite for setting stop-label to $nd$ is to connect all maximal bicliques sharing $nd.R_U$ and $nd.R_V$ into the same partial-\BCPC. The stop-label of $ nd $ is set to one of the maximal bicliques in that partial-\BCPC
\end{enumerate}
We now describe how stop-labels contribute to pruning in identifying branches.

\begin{enumerate}
	\item For a descendant node $ nd $ of an $(\alpha, \beta)$-node, before setting a stop-label to $nd$, it is sufficient to merge the stop-labels of all its child nodes (also real nodes) into the same partial-\BCPC using union-find set. This eliminates the need to traverse all maximal biclique branches within the subtree rooted at $nd$.
	\item For $(\alpha, \beta)$-node $ nd $, in identifying all maximal biclique branches that share $nd.R_U$ and $nd.R_V$, the search terminates as soon as it encounters a node with a stop-label and directly adds the stop-label to the set $\PPr$.
\end{enumerate}

\begin{example}\label{exm:stoplabel}
	Let $ \alpha=1,\beta=1 $, when processing \( nd \) in Figure~\ref{fig:example_beyond_stoplabel} (b):
	
	During the search of subtree rooted at \( nd \), the process terminates upon encountering the stop-label of child node of $ nd $. This stop-label is \( B_4 \), with \( B_5 \) belonging to the same partial-\BCPC as \( B_4 \).
	
	During the search beyond the subtree rooted at \( nd \), the search process terminates upon detecting the stop-label of \( nd' \).
	
	Finally, the two stop-labels (\( B_1 \) and \( B_4 \)) are connected; concurrently, the partial-\BCPCs they belong to are merged into a larger partial-\BCPC.
\end{example}

\begin{algorithm}[t]
	\scriptsize
	\caption{Partial-\BCPC based solution with optimizations}
	\label{alg:bcpc_par_bcpc_slabel}
	\KwIn{Bipartite graph $ G=(U,V,E) $; Two integer $ \alpha,\beta $}
	\KwOut{All \BCPC in $ G $}
	Lines~1-3 of Algorithm~\ref{alg:bcpc_par_bcpc_basic}\;
	Let $\PPr \gets \emptyset$, $ root\gets $ root node of $ T $\;
	Let $ ST(nd),nd\in N $ be the stop-label of $ nd $\;
	\tcc{compute partial-\BCPCs}
	\kw{Postorder}$ (root) $\;

	\tcc{traverse \MBAG with partial-\BCPCs}
	Lines~9-11 of Algorithm~\ref{alg:bcpc_par_bcpc_basic}\;
	
	\Return $ UF $;

	\vspace*{0.1cm}
	{\bf Procedure} {\kw{Postorder}$ (node) $}
	
	\lForEach{child of $ node $, $ node' $, in creation order of $ node' $}{
		\kw{Postorder}$ (node') $
	}
	\If{$ node $ is an $ (\alpha,\beta) $-node}{
		\kw{ProcessNode+}$ (root,node.RCX_U\cup node.RCX_V,node.R_U\cup node.R_V) $;
	}
	\ElseIf{$ node $ is a real node and a descendant of an $ (\alpha,\beta) $-node}{
		\kw{ProcessNode+}$ (node) $;
	}

	\If{$ \PPr $ is not empty}{
		\lFor{$ B\in \PPr $}{$ UF.union(B,\PPr[0]) $}
		$ ST(node)\gets \PPr[0] $, $ \PPr\gets \emptyset $;
	}

	\vspace*{0.1cm}
	{\bf Procedure} {\kw{ProcessNode+}$ (node,UV,RUV) $}
	
	\ForEach{child of $ node $, $ node' $, in creation order of $ node' $}{
		Let $ (node,node',u)\in \vec{E} $\;
		\If{$ node' $ is a real node}{
			\If{$ u\in UV $}{
				\lIf{$ ST(node')=None $}{
					\kw{ProcessNode+}$ (node',UV,RUV) $
				}
				\lElse{
					$ \PPr\gets \PPr\cup\{ST(node')\} $
				}
			}
		}
		
		\lIf{$ u\in RUV $}{\textbf{break}}
	}

	\vspace*{0.1cm}
	{\bf Procedure} {\kw{ProcessNode+}$ (node) $}
	
	\lIf{$ node $ is a leaf node}{$ \PPr\gets \PPr\cup \{(node.R_U,node.R_V)\}$}
	\Else{
		\ForEach{$ node' $, $ node' $ is a child of $ node $ and a real node}{
			$ \PPr\gets \PPr\cup \{ST(node')\} $
		}
	}
	
\end{algorithm}

\stitle{Implementation.} Based on the above strategies, we revise the original framework and present Algorithm~\ref{alg:bcpc_par_bcpc_slabel}. Algorithm~\ref{alg:bcpc_par_bcpc_slabel} performs an initialization step in Line~1 that is similar to that of Algorithm~\ref{alg:bcpc_par_bcpc_basic}. In Line~3, $ ST(nd) $ denotes the stop-label of $ nd $. The core component of Algorithm~\ref{alg:bcpc_par_bcpc_slabel} is the \kw{Postorder} function (Lines~4 and~7). This function performs a post-order traversal of the \MBE tree, a process that can be executed during the \MBE progress. The \kw{Postorder} function is responsible for handling both the $(\alpha,\beta)$-node itself—by invoking \kw{ProcessNode+}$ (node, UV, RUV) $ in Line~10—and the real nodes among its descendant nodes—by invoking \kw{ProcessNode+}$ (node) $ in Line~12. Both processing routines may add certain maximal bicliques to $\PPr$. Finally, in Lines~13–15, Algorithm~\ref{alg:bcpc_par_bcpc_slabel} connects all the maximal bicliques in $\PPr$ into the same partial-\BCPC, and it set the first biclique in $ \PPr $ as the stop-label of $ node $ (Line~15). Similar to Algorithm~\ref{alg:bcpc_par_bcpc_basic}, Algorithm~\ref{alg:bcpc_par_bcpc_slabel} traverses the \MBAG based on partial-\BCPCs to obtain the final \BCPC (Line~5).

When processing an $(\alpha,\beta)$-node (Line~10), the \kw{ProcessNode+} function starts traversing the \MBE tree from its root node. It mainly explores real nodes (Line~19), since any branch containing a maximal biclique must be composed of real nodes. Moreover, any vertex $u$ along the traversal path must belong to the set $UV$ (Line~20). The recursive calls end immediately upon encountering a node that has a stop-label assigned (Lines~22). After processing a child node $ node' $, if the vertex $u$ along the traversal path belongs to the set $RUV$, there is no need to explore the remaining child nodes and their corresponding branches (Line~23). This is because, according to the enumeration principle of the \MBE algorithm, any maximal biclique that includes $RUV$ can not be found in those subsequent branches. Since the goal of this function is to identify maximal bicliques that contain $RUV$, further traversal becomes unnecessary in this case.

When processing the descendant nodes of an $(\alpha,\beta)$-node (Line~12), if the parameter $ node $ is already a leaf node, the maximal biclique it represents can be directly added to $\PPr$. Otherwise, the stop-labels of all its child nodes are added to $\PPr$.

\stitle{Analysis.} We now proceed to analyze the correctness of Algorithm~\ref{alg:bcpc_par_bcpc_slabel}. The key lies in demonstrating that the partial-\BCPCs computed in \kw{Postorder} (Line~3) are correct.

\begin{theorem}\label{thm:postorder_correct}
	For each $ (\alpha,\beta) $-node processed in Algorithm~\ref{alg:bcpc_par_bcpc_slabel} (Line~9), \kw{ProcessNode+} connects all maximal bicliques in the bipartite graph that share \( (node.R_U, node.R_V) \) into the same partial-\BCPC.
\end{theorem}
\ifdefined\showproof
\begin{proof}
	Consider a sequence of \( (\alpha, \beta) \)-nodes \( (nd_1, nd_2, \ldots, nd_n) \), which follow the processing order in \kw{Postorder} (i.e., post-order traversal). We first prove that during the processing of \( nd_1 \), \kw{ProcessNode+} connects all maximal bicliques that share \( (nd_1.R_U, nd_1.R_V) \) into the same partial-\BCPC. Since \( nd_1 \) is the first \( (\alpha, \beta) \)-node, it must also be in the first enumerated maximal biclique branch (if $ nd_1 $ appears in an earlier branch, then that earlier or an even earlier branch is likely the first one containing a maximal biclique; if $ nd_1 $ appears in a later branch, then it can not be the first $ (\alpha,\beta) $-node). Therefore, all maximal biclique branches that contain \( (nd_1.R_U, nd_1.R_V) \) must lie in the subtree rooted at \( nd_1 \), and can not appear in any subtree processed before \( nd_1 \).
	
	Using mathematical induction, assume that \kw{ProcessNode+} connects all maximal bicliques that share $ (nd_i.R_U, nd_i.R_V) $ when processing $ nd_1, nd_2, ..., nd_i $. Next, we prove the case for $ i+1 $. For $ nd_{i+1} $, let $ R_U = nd_{i+1}.R_U $ and $ R_V = nd_{i+1}.R_V $. According to the \MBE process \cite{dai2023hereditary}, the branches containing maximal bicliques with $ R_U $ and $ R_V $ are either generated before $ nd_{i+1} $, or located in the subtree rooted at $ nd_{i+1} $. For the maximal biclique branches generated before $ nd_{i+1} $, their corresponding $ (\alpha, \beta) $-node $ nd_j $ must satisfy $ j < i+1 $, so we only need to collect $ ST(nd_j) $, as it represents the partial-\BCPCs of those maximal bicliques. For the maximal bicliques in the subtree of $ nd_{i+1} $, we only need to collect the stop-labels of the child nodes of $ nd_{i+1} $, because the maximal bicliques under each child node have already been connected into partial-\BCPCs (since the traversal is post-order). Therefore, the case for $ nd_{i+1} $ is proved.
\end{proof}
\else
\fi

\begin{theorem}\label{thm:pbcpc_op_correct}
	Algorithm~\ref{alg:bcpc_par_bcpc_slabel} correctly computes all \BCPCs.
\end{theorem}
\ifdefined\showproof
\begin{proof}
	The proof of this theorem is similar to that of Theorem~\ref{thm:pbcpc_basic_correct}. Since Theorem~\ref{thm:postorder_correct} proves that the partial-\BCPCs in Algorithm~\ref{alg:bcpc_par_bcpc_slabel} are obtained by connecting maximal bicliques that share the $R_U$ and $R_V$ of $(\alpha, \beta)$-nodes, these partial-\BCPCs are correct. Moreover, the process of computing the final \BCPCs in Algorithm~\ref{alg:bcpc_par_bcpc_slabel} (Line~5) is the same as in Algorithm~\ref{alg:bcpc_par_bcpc_basic}, thus, Algorithm~\ref{alg:bcpc_par_bcpc_slabel} can correctly compute the \BCPCs.
\end{proof}
\else
\fi

It should be noted that the partial-\BCPCs computed by Algorithm~\ref{alg:bcpc_par_bcpc_slabel} are not the same as those computed by Algorithm~\ref{alg:bcpc_par_bcpc_basic}. We illustrate this with the following theorem.

\begin{theorem}\label{thm:pbcpc_num_op}
	Given a bipartite graph $ G $, two integer $ \alpha,\beta $, let $ pbcpc $ be the number of partial-\BCPCs computed by Algorithm~\ref{alg:bcpc_par_bcpc_basic}, $ pbcpc+ $ be the number of partial-\BCPCs computed by Algorithm~\ref{alg:bcpc_par_bcpc_slabel}, $ bcpc $ be the number of \BCPCs, the following inequality holds: $ pbcpc\geq pbcpc+\geq bcpc $.
\end{theorem}
\ifdefined\showproof
\begin{proof}
	We only need to prove $ pbcpc\geq pbcpc+ $. When processing an $(\alpha, \beta)$-node, Algorithm~\ref{alg:bcpc_par_bcpc_slabel} not only connects maximal bicliques in the subtree rooted at that node, but also further connects maximal bicliques that were enumerated in earlier branches and share $ R_U,R_V $ of the $(\alpha, \beta)$-node. Moreover, the \kw{ProcessNode+} procedure in Algorithm~\ref{alg:bcpc_par_bcpc_slabel} also processes $(\alpha, \beta)$-nodes that belong to virtual nodes. Therefore, $ pbcpc\geq pbcpc+ $.
\end{proof}
\else
\fi

\begin{theorem}\label{thm:pbcpc_op_complexity}
	The time and space complexity of Algorithm~\ref{alg:bcpc_par_bcpc_slabel} are $ O((1-\frac{1}{pbcpc+})n_{biclique}^2n) $ and $ O(n_{biclique}n) $ respectively, where $ n=|U|+|V| $, $ pbcpc+ $ is the number of partial-\BCPCs, $ n_{biclique} $ is the number of maximal bicliques.
\end{theorem}
\ifdefined\showproof
\begin{proof}
	Let $ m=|E| $, the time complexity of maximal biclique enumeration is $ O(2^{n/2}m) $ \cite{dai2023hereditary}. Unlike Algorithm~\ref{alg:bcpc_par_bcpc_basic}, Algorithm~\ref{alg:bcpc_par_bcpc_slabel} handles not only $ (\alpha, \beta) $-nodes (Lines~9-10) but also their child nodes (Lines~11-12). For processing the child nodes of $ (\alpha, \beta) $-nodes, Algorithm~\ref{alg:bcpc_par_bcpc_slabel} only needs to traverse one layer of their child nodes, resulting in an overall time complexity not exceeding $ O(2^{n/2}m) $. Regarding the processing of $ (\alpha, \beta) $-nodes, let $ x $ be the total number of such nodes (including both real nodes and virtual nodes), $ d $ be the maximum depth of real $ (\alpha,\beta) $-nodes. When processing each of these nodes (Line~10), the procedure \kw{ProcessNode+} will access at most all real $ (\alpha, \beta) $-nodes. Since the number of child nodes of each node in the \MBE tree does not exceed $ n $, the total time complexity of the above process is $ O(xn^d) = O(2^{n/2}mn^d) $. The time complexity analysis for the traversal of \MBAG in Algorithm~\ref{alg:bcpc_par_bcpc_slabel} is similar to that of Algorithm~\ref{alg:bcpc_par_bcpc_basic}. As a result, the time complexity of Algorithm~\ref{alg:bcpc_par_bcpc_slabel} is $ O(2^{n/2}mn^{d}+(1-\frac{1}{pbcpc+})n_{biclique}^2n) $. Since $ O(n_{biclique})=O(2^{n/2}) $ \cite{prisner2000bicliques,dai2023hereditary}, the time complexity can be rewritten as $ O((1-\frac{1}{pbcpc+})n_{biclique}^2n) $.
	
	Similar to Algorithm~\ref{alg:bcpc_par_bcpc_basic}, the space complexity of Algorithm~\ref{alg:bcpc_par_bcpc_slabel} is also $ n_{biclique}n $.
\end{proof}
\else
\fi

\stitle{Discussion.} The inequality $pbcpc\geq pbcpc+$ in Theorem~\ref{thm:pbcpc_num_op} forms the basis for Algorithm~\ref{alg:bcpc_par_bcpc_slabel} outperforming Algorithm~\ref{alg:bcpc_par_bcpc_basic}, since it implies that Algorithm~\ref{alg:bcpc_par_bcpc_slabel} can traverse a smaller \MBAG. Experiments (Figure~\ref{fig:pbcpc_num}) show that $ pbcpc+ $ is significantly smaller than $ pbcpc $—for example, smaller by an order of magnitude, and Algorithm~\ref{alg:bcpc_par_bcpc_slabel} can be an order of magnitude faster than Algorithm~\ref{alg:bcpc_par_bcpc_basic} (Table~\ref{tab:all_performance}).

\section{Novel $ (\alpha,\beta) $-Biclique based Solution}\label{sec:biclique-list}
Existing \BCPC detection methods are all based on the \MBAG approach. In the previous section, we reduce the \MBAG using partial-\BCPCs, but it is still essentially based on \MBAG and remains inefficient for certain datasets where \MBAG is extremely large. This section introduces a novel \BCPC detection framework, which can also leverage the partial-\BCPCs proposed in Section~\ref{sec:new_mbag} to accelerate the detection process. We first introduce this framework.

\subsection{The Basic Framework}

\stitle{Key idea.} Note that in the previous section, each edge in the \MBAG essentially represents an $(a,b)$-biclique, where $a \geq \alpha$ and $b \geq \beta$. A straightforward idea is to enumerate all $(\alpha,\beta)$-bicliques in the bipartite graph, and for each enumerated $(\alpha,\beta)$-biclique $B$, connect all maximal bicliques containing $B$ using union-find set.

\begin{algorithm}[t]
	\scriptsize
	\caption{$ (\alpha,\beta) $-Biclique based solution}
	\label{alg:bcpc_ablist_frame}
	\KwIn{Bipartite graph $ G=(U,V,E) $; Two integer $ \alpha,\beta $}
	\KwOut{All \BCPC in $ G $}
	
	Let $ UF $ be an initial union-find set\;
	$ \Br\gets $ set of all maximal bicliques in $ G $\;
	$ \Br\gets \{B|B\in \Br,|B.X|\geq \alpha,|B.Y|\geq \beta\} $\;
	
	Construct 2-hop graph $ \vec{H}=(U,\vec{E_h}) $ on vertex set $ U $ \cite{yang2023p}; \tcc{$ \vec{H}=(U,\vec{E_h}) $: a direct graph, $ (u,v)\in \vec{E_h} $ indicates that $ u<v $, $ u,v $ have common neighbors in $ V $}
	
	\kw{BicliqueListing}$ (\vec{H},\emptyset,V) $\;
	\Return $ UF $;
	
	\vspace*{0.1cm}
	{\bf Procedure} {\kw{BicliqueListing}$ (\vec{H},R_U,C_V) $}
	
	\If{$ |R_U|=\alpha $}{
		\lForEach{$ R_V\subseteq C_V,|R_V|=\beta $}{
			Connect maximal bicliques sharing $ (R_U,R_V) $ in $ \Br $ with $ UF.union(\ast) $
		}
		\Return;
	}
	
	\ForEach{$ u\in \vec{H} $}{
		$ C_V'\gets C_V\cap Nei_G(u) $\;
		\lIf{$ |C_V'|<\beta $ or $ Nei_{\vec{H}}(u)<\alpha-|R_U|-1 $}{\textbf{continue}}
		
		Let $ \vec{H}' $ be induced subgraph of $ \vec{H} $ on $ Nei_{\vec{H}}(u) $\;
		\kw{BicliqueListing}$ (\vec{H}',R_U\cup \{u\},C_V') $;
	}
	
\end{algorithm}

\stitle{Implementation.} Algorithm~\ref{alg:bcpc_ablist_frame} presents the basic framework of the $(\alpha, \beta)$-biclique based solution. It is built upon the $(\alpha, \beta)$-biclique enumeration framework proposed in \cite{yang2023p}, where the core idea is to enumerate combinations of vertices in the $U$ set while maintaining their common neighbors in the $V$ set. This framework constructs an ordinary directed graph, called the \textit{2-hop graph}, on the vertices of the $U$ set to accelerate the enumeration process.

In Lines~2–3, Algorithm~\ref{alg:bcpc_ablist_frame} prepares the maximal bicliques that need to be connected. In Line~4, it constructs the 2-hop graph according to \cite{yang2023p}. Then, in Line~5, it executes the $(\alpha, \beta)$-biclique enumeration framework. Within \kw{BicliqueListing}, for each enumerated $(\alpha, \beta)$-biclique (Line~9), all maximal bicliques that share this biclique are connected together.

\stitle{Analysis.} Now we proceed to analyze the correctness and complexity of Algorithm~\ref{alg:bcpc_ablist_frame}.

\begin{theorem}\label{thm:bcpc_ablist_frame_correct}
	Algorithm~\ref{alg:bcpc_ablist_frame} correctly computes all \BCPCs.
\end{theorem}
\ifdefined\showproof
\begin{proof}
	This theorem can be easily proven as follows: Algorithm~\ref{alg:bcpc_ablist_frame} is capable of enumerating all $ (\alpha, \beta) $-bicliques and connecting all maximal bicliques that share these bicliques. Consequently, the final \BCPCs are correct.
\end{proof}
\else
\fi

\begin{theorem}\label{thm:bcpc_ablist_frame_complexity}
	The time and space complexity of Algorithm~\ref{alg:bcpc_ablist_frame} are $ O(a(\vec{H})^{\alpha-2}|\vec{E_h}||V|+\Delta n_{biclique}) $ and $ O(n_{biclique}n) $ respectively, where $ n=|U|+|V| $, $ a(\vec{H}) $ is the arboricity of $\vec{H}$ \cite{yang2023p}, $ \Delta $ is the number of $ (\alpha, \beta) $-bicliques, $ n_{biclique} $ is the number of maximal bicliques.
\end{theorem}

\ifdefined\showproof
\begin{proof}
	The time complexity of maximal biclique enumeration is $T_{biclique}=2^{n/2}|E| $ \cite{dai2023hereditary}. The time complexity of enumerating $ (\alpha, \beta) $-biclique is $ O(a(\vec{H})^{\alpha-2}|\vec{E_h}||V|+\Delta) $ \cite{yang2023p}. Algorithm~\ref{alg:bcpc_ablist_frame} connects all maximal bicliques sharing an $ (\alpha, \beta) $-biclique (Line~9), thus, the time complexity of Algorithm~\ref{alg:bcpc_ablist_frame} is $ O(T_{biclique}+a(\vec{H})^{\alpha-2}|\vec{E_h}||V|+\Delta n_{biclique}) $. Since $ O(n_{biclique})=O(2^{n/2}) $ \cite{prisner2000bicliques,dai2023hereditary}, $ O(\Delta)=O(n^{\alpha+\beta}) $ \cite{yang2023p}, the time complexity can be rewritten as $ O(a(\vec{H})^{\alpha-2}|\vec{E_h}||V|+\Delta n_{biclique}) $.
	
	Algorithm~\ref{alg:bcpc_ablist_frame} does not store all $ (\alpha, \beta) $-bicliques, thus, the space complexity of Algorithm~\ref{alg:bcpc_ablist_frame} is $ O(n_{biclique}n) $.
\end{proof}
\else
\fi

\subsection{Pruning with Maximal Bicliques and Partial-\BCPCs}

\stitle{Challenges in the basic framework.} It is evident that the basic framework in Algorithm~\ref{alg:bcpc_ablist_frame} is highly inefficient, as the number of $(\alpha, \beta)$-bicliques grows exponentially with respect to $\alpha$ and $\beta$ \cite{yang2023p}. Therefore, enumerating all $(\alpha, \beta)$-bicliques and connecting the corresponding maximal bicliques is impractical.

\stitle{Pruning with Maximal Bicliques.} The key to addressing the above challenge is to reduce the search space in the enumeration process of $(\alpha, \beta)$-bicliques — that is, to prune in advance those branches that do not contribute to connecting multiple maximal bicliques.

\begin{theorem}\label{thm:prune_biclique}
	Given a bipartite graph $ G $, for a biclique $B$, if the number of maximal bicliques sharing it is 0, or all those maximal bicliques are in the same \BCPC, then for any other biclique $B'$ such that $B \subseteq B'$, the number of maximal bicliques sharing $B'$ is also 0, or they are in the same \BCPC.
\end{theorem}
\ifdefined\showproof
\begin{proof}
	It is easy to see that any maximal biclique containing $B'$ must also contain $B$, so the set of maximal bicliques containing $B'$ is a subset of those containing $B$. Therefore, Theorem~\ref{thm:prune_biclique} can be easily proved.
\end{proof}
\else
\fi

Theorem~\ref{thm:prune_biclique} reveals the core idea of pruning the $(\alpha, \beta)$-biclique enumeration space: during the enumeration process, for each intermediate small biclique generated, we can maintain the set of maximal bicliques containing it. If this set is empty or all the maximal bicliques in the set already belong to the same \BCPC, then the current enumeration branch can be safely pruned.

\begin{algorithm}[t]
	\scriptsize
	\caption{$ (\alpha,\beta) $-Biclique based solution with maximal biclique}
	\label{alg:bcpc_ablist_mbiclique}
	\KwIn{Bipartite graph $ G=(U,V,E) $; Two integer $ \alpha,\beta $}
	\KwOut{All \BCPC in $ G $}
	
	Let $ UF $ be an initial union-find set\;
	$ \Br\gets $ set of all maximal bicliques in $ G $\;
	$ \Br\gets \{B|B\in \Br,|B.X|\geq \alpha,|B.Y|\geq \beta\} $\;

	Construct 2-hop graph $ \vec{H}=(U,\vec{E_h}) $ on vertex set $ U $ \cite{yang2023p}; \tcc{$ \vec{H}=(U,\vec{E_h}) $: a direct graph, $ (u,v)\in \vec{E_h} $ indicates that $ u<v $, $ u,v $ have common neighbors in $ V $}
	
	\kw{BicliqueListing+}$ (\vec{H},\emptyset,V) $\;
	\Return $ UF $;
	
	\vspace*{0.1cm}
	{\bf Procedure} {\kw{BicliqueListing+}$ (\vec{H},R_U,C_V) $}

	$ RB\gets \{B|R_U\subseteq B.X,B\in \Br\} $; \tcc{maximal bicliques sharing $ R_U $}
	\textbf{if} $ |\{UF.find(B)|B\in RB\}|\leq 1 $ \textbf{then return;} \tcc{there is only 1 or 0 $ UF $ disjoint set in $ RB $}
	
	\If{$ |R_U|=\alpha $}{
		\ForEach{$ R_V\subseteq C_V,|R_V|=\beta $}{
			$ RB\gets \{B|R_V\subseteq B.Y,B\in RB\} $\;
			\textbf{if} $ |\{UF.find(B)|B\in RB\}|\leq 1 $ \textbf{then continue;} \tcc{there is only 1 or 0 $ UF $ disjoint set in $ RB $}
			
			Connect maximal bicliques in $ RB $ with $ UF.union(\ast) $;
		}
		\Return;
	}
	
	\ForEach{$ u\in \vec{H} $}{
		$ C_V'\gets C_V\cap Nei_G(u) $\;
		\lIf{$ |C_V'|<\beta $ or $ Nei_{\vec{H}}(u)<\alpha-|R_U|-1 $}{\textbf{continue}}
		
		Let $ \vec{H}' $ be induced subgraph of $ \vec{H} $ on $ Nei_{\vec{H}}(u) $\;
		\kw{BicliqueListing+}$ (\vec{H}',R_U\cup \{u\},C_V') $;
	}
	
\end{algorithm}

\stitle{Implementation.} Algorithm~\ref{alg:bcpc_ablist_mbiclique} presents the implementation of pruning the $(\alpha, \beta)$-biclique enumeration process using maximal bicliques. The difference between Algorithm~\ref{alg:bcpc_ablist_mbiclique} and Algorithm~\ref{alg:bcpc_ablist_frame} lies in the \kw{BicliqueListing+} function. Specifically, in Line~8, \kw{BicliqueListing+} collects maximal bicliques that share $(R_U, \emptyset)$. In Line~9, if the maximal bicliques in $RB$ belong to only one or zero union-find sets, then the entire enumeration branch can be pruned. In Lines~12–13, the algorithm further checks whether the maximal bicliques in $RB$ that share $R_V$ belong to more than one union-find set. If so, it connects them using $ UF.union(\ast) $.

\stitle{Analysis.} Now we proceed to analyze the correctness and complexity of Algorithm~\ref{alg:bcpc_ablist_mbiclique}.

\begin{theorem}\label{thm:bcpc_ablist_mbiclique_correct}
	Algorithm~\ref{alg:bcpc_ablist_mbiclique} correctly computes all \BCPCs.
\end{theorem}
\ifdefined\showproof
\begin{proof}
	Compared with Algorithm~\ref{alg:bcpc_ablist_frame}, Algorithm~\ref{alg:bcpc_ablist_mbiclique} prunes the branches that can not connect multiple $ UF $ sets based on Theorem~\ref{thm:prune_biclique}. Therefore, Algorithm~\ref{alg:bcpc_ablist_mbiclique} maintains the same correctness as Algorithm~\ref{alg:bcpc_ablist_frame}.
\end{proof}
\else
\fi

\begin{theorem}\label{thm:bcpc_ablist_mbiclique_complexity}
	The time and space complexity of Algorithm~\ref{alg:bcpc_ablist_mbiclique} are $ O((a(\vec{H}_m)^{\alpha-2}|\vec{E}_{hm}||V_m|+\Delta_m) n_{biclique}) $ and $ O(n_{biclique}n) $ respectively. $ n=|U|+|V|$. Let \( U_m \) and \( V_m \) be the vertex sets shared by any two maximal bicliques, \( G_m \) be the induced subgraph on \( U_m \) and \( V_m \), and \( \vec{H}_m = (U_m, \vec{E}_{hm}) \) be the 2-hop graph based on \( G_m \). $ a(\vec{H}_m) $ is the arboricity of $\vec{H}_m$ \cite{yang2023p}, $ \Delta_m $ is the number of $ (\alpha, \beta) $-bicliques in \( G_m \), $ n_{biclique} $ is the number of maximal bicliques.
\end{theorem}
\ifdefined\showproof
\begin{proof}
	The time complexity of maximal biclique enumeration is $T_{biclique}=2^{n/2}|E| $ \cite{dai2023hereditary}. Compared with Algorithm~\ref{alg:bcpc_ablist_frame}, Algorithm~\ref{alg:bcpc_ablist_mbiclique} prunes the branches that can not connect multiple $ UF $ sets based on Theorem~\ref{thm:prune_biclique}. This implies that vertices initially contained in only one maximal biclique need not be considered during the $ (\alpha, \beta) $-biclique enumeration process. Therefore, Algorithm~\ref{alg:bcpc_ablist_mbiclique} can be regarded as performing $ (\alpha, \beta) $-biclique enumeration on \( G_m \). Considering Lines~8-9 and Lines~12-13 of Algorithm~\ref{alg:bcpc_ablist_mbiclique}, its time complexity is $ O(T_{biclique}+(a(\vec{H}_m)^{\alpha-2}|\vec{E}_{hm}||V_m|+\Delta_m) n_{biclique}) $. Since $ O(n_{biclique})=O(2^{n/2}) $ \cite{prisner2000bicliques,dai2023hereditary}, $ O(\Delta_m)=O((U_m+V_m)^{\alpha+\beta}) $ \cite{yang2023p}, the time complexity can be rewritten as $ O((a(\vec{H}_m)^{\alpha-2}|\vec{E}_{hm}||V_m|+\Delta_m) n_{biclique}) $.
	
	Similarly, the space complexity of Algorithm~\ref{alg:bcpc_ablist_mbiclique} is still $ O(n_{biclique}n) $.
\end{proof}
\else
\fi

\stitle{Pruning with Partial-\BCPC.} As shown in Section~\ref{sec:new_mbag}, a partial-\BCPC is an incomplete \BCPC. Based on the partial-\BCPCs, we can determine in advance that certain maximal bicliques belong to the same \BCPC. Therefore, they enable the $(\alpha, \beta)$-biclique enumeration process in Algorithm~\ref{alg:bcpc_ablist_mbiclique} to reach pruning conditions earlier.

\begin{algorithm}[t]
	\scriptsize
	\caption{$ (\alpha,\beta) $-Biclique based solution with partial-\BCPC}
	\label{alg:bcpc_ablist_pbcpc}
	\KwIn{Bipartite graph $ G=(U,V,E) $; Two integer $ \alpha,\beta $}
	\KwOut{All \BCPC in $ G $}
	
	Lines~1-3 of Algorithm~\ref{alg:bcpc_par_bcpc_slabel}; \tcc{partial-\BCPC computation}
	
	Lines~2-5 of Algorithm~\ref{alg:bcpc_ablist_mbiclique}; \tcc{connect maximal bicliques by $ (\alpha,\beta) $-biclique listing}
	
	\Return $ UF $;
	
%
%
%
%
%
	
\end{algorithm}

\stitle{Implementation.} Algorithm~\ref{alg:bcpc_ablist_pbcpc} is the algorithm that leverages partial-\BCPCs, and it consists of two main parts. First, it computes the partial-\BCPC (Line~1), which are recorded in the union-find structure ($ UF $). Then, in Line~2, these partial-\BCPCs are used to further prune the $(\alpha, \beta)$-biclique listing process.

\stitle{Analysis.} Now we proceed to analyze the correctness and complexity of Algorithm~\ref{alg:bcpc_ablist_pbcpc}.

\begin{theorem}\label{thm:bcpc_ablist_pbcpc_correct}
	Algorithm~\ref{alg:bcpc_ablist_pbcpc} correctly computes all \BCPCs.
\end{theorem}

\begin{theorem}\label{thm:bcpc_ablist_pbcpc_complexity}
	The time and space complexity of Algorithm~\ref{alg:bcpc_ablist_pbcpc} are $ O((a(\vec{H}_p)^{\alpha-2}|\vec{E}_{hp}||V_p|+\Delta_p) n_{biclique}) $ and $ O(n_{biclique}n) $ respectively. $ n=|U|+|V| $. Let \( U_p \) and \( V_p \) be the vertex sets shared by any two partial-\BCPCs, \( G_p \) be the induced subgraph on \( U_p \) and \( V_p \), and \( \vec{H}_p = (U_p, \vec{E}_{hp}) \) be the 2-hop graph based on \( G_p \). $ a(\vec{H}_p) $ is the arboricity of $\vec{H}_p$ \cite{yang2023p}, $ \Delta_p $ is the number of $ (\alpha, \beta) $-bicliques in \( G_p \), $ n_{biclique} $ is the number of maximal bicliques.
\end{theorem}
\ifdefined\showproof
The proves of Theorem~\ref{thm:bcpc_ablist_pbcpc_correct} and Theorem~\ref{thm:bcpc_ablist_pbcpc_complexity} are similar to those of Theorem~\ref{thm:bcpc_ablist_mbiclique_correct} and Theorem~\ref{thm:bcpc_ablist_mbiclique_complexity}.
\else
\fi

\section{Experiments}\label{sec:experiment}

\subsection{Experimental Setup}
\stitle{Datasets.} We use 10 real-world bipartite graph datasets, whose detailed information is presented in Table~\ref{tab:datasets}. All the datasets are downloaded from \url{http://konect.cc/}.

\begin{table}[t]\vspace*{0.1cm}
	\small
	\centering
	\caption{Datasets, $ \overline{d}_U $ is the average degree of vertices in $ U $, $ \overline{d}_V $ is the average degree of vertices in $ V $, $N_m$ is the number of maximal bicliques, $N_o$ is the number of edges in maximal biclique adjacent graph, K=$10^3$, M=$10^6$, B=$10^9$, `-' means timeout} \vspace*{-0.2cm} \label{tab:datasets}
	\resizebox{0.99\columnwidth}{!}{
		\begin{tabular}{|c | c | c | c | c|c|c|c|}
			\hline
			\rule{0pt}{9pt}Datasets &  $ |U| $ & $ |V| $ & $ |E| $ & $\overline{d}_U$& $\overline{d}_V$&$N_m$&$N_o$ \cr 
			\hline
			\kw{Youtube}&94,238&30,087&293,360&3.11&9.75&1,769,331&2B\cr
			\kw{Bookcrossing}&77,802&185,955&433,652&5.57&2.33&155,391&1B \cr
			\kw{Github}&56,519&120,867&440,237&7.79&3.64&55,260,550&- \cr
			\kw{Citeseer}&105,353&181,395&512,267&4.86&2.82&54,083&962K \cr
			\kw{Stackoverflow}&545,195&96,678&1,301,942&2.39&13.47&2,922,148&26B \cr
			\kw{Twitter}&175,214&530,418&1,890,661&10.79&3.56&5,102,542&13B\cr
			\kw{Imdb}&685,568&186,414&2,715,604&3.96&14.57&1,809,175&1B \cr
			\kw{Actor2}&303,617&896,302&3,782,463&12.46&4.22&3,761,666&2B \cr
			\kw{Amazon}&2,146,057&1,230,915&5,743,258&2.68&4.67&5,702,211&108B \cr
			\kw{DBLP}&1,953,085&5,624,219&12,282,059&6.29&2.18&1,281,831&85M \cr
			\hline
	\end{tabular}}
	\vspace*{-0.5cm}
\end{table}

\begin{table*}[t]\vspace*{0.1cm}
	\small
	\centering
	\setlength{\tabcolsep}{2.8pt}
	\caption{Performance (s) of all algorithms on all datasets. The best and sub-best results are bold and underlined respectively. ``-'' denotes out of time ($ \geq 7$ days)} \vspace*{-0.2cm} \label{tab:all_performance}
	\resizebox{1.0\textwidth}{!}{
		\begin{tabular}{|c | c | c | c | c|c|c|c|c|c|c|c|c|c|c|c|c|c|c|c|c|c|}
			\hline
			\multicolumn{2}{|c|}{Datasets} &  \multicolumn{4}{c|}{\kw{Youtube}} & \multicolumn{4}{c|}{\kw{Bookcrossing}} & \multicolumn{4}{c|}{\kw{Github}} & \multicolumn{4}{c|}{\kw{Citeseer}}&\multicolumn{4}{c|}{\kw{Stackoverflow}} \cr 
			\hline
			$\alpha$&\diagbox{Algo}{$\beta$}&2&4&6&8&2&4&6&8&2&4&6&8&2&4&6&8&2&4&6&8 \cr
			\hline
			\multirow{6}{*}{2}&\MBAG&13,376 & 11,492 & 6,072 & 1,902 &35 &2 &\underline{0.9} & \underline{0.7}&- &- &- &- &1 & 0.8 & 0.7 & 0.6 &32,598 & 17,967 & 4,287 & 844 \cr
			&\pbcpc&4,651 & 7,283 & 6,942 & 3,207 &42 & 3 & 1 & 1 &- &- &- &- &1 & \underline{0.8} & \underline{0.7} & 0.6 &30,051 & 25,220 & 8,811 & 2,177\cr
			&\pbcpcp & 197 & 1,914 & 2,787 & 1,484 &20 & 3 & 1 & 0.8 &44,276&-&-&-&1 & 0.9 & 0.8 & 0.6 &2,564 & 7,144 & 3,467 & 987\cr
			&\biclique& 135 & 358 & 2,995 & - & 3 & 26 & 93,773 & - &- &- &- &- &1 & 1 & 64 & 6,999 & 2,038 & 3,118 & - & -  \cr
			&\bicliquem& \underline{59} & \underline{70} & \underline{43} & \underline{22} &\textbf{2} & \textbf{1} & \textbf{0.8}  & \textbf{0.6}  &\underline{24,581} &- &- &- &\underline{1} & \textbf{0.8} & \textbf{0.7} & \underline{0.6} &\underline{486} & \underline{208} & \textbf{119} & \textbf{80}\cr
			&\bicliquep& \textbf{34} & \textbf{43} & \textbf{31} & \textbf{20} &\underline{3} & \underline{2} & 1 & 1 &\textbf{7,718} & \textbf{34,478} & \textbf{33,948} & \textbf{14,272} &\textbf{1} & 0.9 & 0.8 & \textbf{0.6} &\textbf{339} & \textbf{159} & \underline{121} & \underline{103}\cr
			\hline
			\multirow{6}{*}{4}&\MBAG&13,113 & 11,358 & 5,941 & 1,793& 3 &\textbf{0.7} & \textbf{0.5} & \textbf{0.4} &- &- &- &- &0.1 & \underline{0.08} & \underline{0.07} & \underline{0.06} &27,582 & 14,496 & 3,308 & 544 \cr
			&\pbcpc&7,748 & 9,568 & 7,802 & 3,197 &4 & 0.9 & 0.6 & 0.5 &- &- &- &- &0.1 & 0.09 & 0.08 & 0.06 &37,632 & 25,336 & 7,496 & 1,476\cr
			&\pbcpcp& 1,450 & 2,964 & 3,404 & 1,618 &4 & 0.8 & 0.6 & 0.5 &- &- &- &- &0.1 & \textbf{0.08} & 0.07 & 0.06 &13,222 & 9,955 & 3,857 & 864\cr
			&\biclique& 3,754 & 5,445 & 5,522 & 2,877 &2 & 1 & 0.8 & 0.6 & - &- &- &- & 0.1 & 0.1 & 0.1 & 0.3 & 1,210 & 5,446 & 748 & 11,595 \cr
			&\bicliquem& \underline{54} & \underline{235} & \underline{140} & \underline{50} &\underline{1} & \underline{0.8}  & 0.6  & 0.5  &- & - & - & \underline{412,886} & \underline{0.1} & 0.09 & \textbf{0.07} & \textbf{0.06} & \underline{306} & \underline{539} & \underline{199} & \underline{56}\cr
			&\bicliquep& \textbf{28} & \textbf{66} & \textbf{49} & \textbf{24} &\textbf{1} & 0.9 & \underline{0.6} & \underline{0.5} &\textbf{11,483} & \textbf{145,121} & \textbf{424,317} & \textbf{157,465} &\textbf{0.1} & 0.09 & 0.08 & 0.07 &\textbf{190} & \textbf{159} & \textbf{76} & \textbf{34}\cr
			\hline
			\multirow{6}{*}{6}&\MBAG&10,936 & 9,409 & 4,635 & 1,312& 1 &\underline{0.6} & 0.5 & 0.4 &- &- &- &- &\textbf{0.03} & \textbf{0.03} & \textbf{0.03} & \textbf{0.03} &9,142 & 4,150 & 561 & 45 \cr
			&\pbcpc&7,736 & 8,370 & 5,855 & 2,080 &1 & 0.7 & 0.5 & 0.4 &- &- &- &- &0.05 & 0.04 & 0.04 & 0.04 &12,482 & 6,893 & 1,134 & 89\cr
			&\pbcpcp& 2,500 & 3,622 & 3,133 & 1,121 &1 & 0.7 & 0.5 & 0.4 &- &- &- &- &0.04 & \underline{0.03} & 0.04 & \underline{0.03} &5,758 & 3,833 & 775 & 76\cr
			&\biclique& - & 3,289 & 34,084 & 3,352 & 36 & 0.9 & 0.5 & 0.4 &- &- &- &- & 0.07 & 0.04 & 0.04 & 0.04 &- & 1,764 & 2,724 & 26   \cr
			&\bicliquem& \underline{44} & \underline{165} & \underline{178} & \underline{35} &\underline{1} & \textbf{0.6}  & \underline{0.5}  & \underline{0.4}  &\underline{13,182} & \underline{444,343} & - & \underline{55,394} & \underline{0.04} & 0.04 & 0.04 & 0.04 &\underline{90} & \underline{131} & \underline{40} & \underline{12}\cr
			&\bicliquep& \textbf{24} & \textbf{57} & \textbf{56} & \textbf{16} &\textbf{1} & 0.7 & \textbf{0.5} & \textbf{0.4} &\textbf{9,450} & \textbf{141,003} & \textbf{158,184} & \textbf{20,785} &0.05 & 0.04 & \underline{0.04} & 0.04 & \textbf{75} & \textbf{72} & \textbf{22} & \textbf{10}\cr
			\hline
			\multirow{6}{*}{8}&\MBAG&7,236& 6,102& 2,832& 744& 1 &\textbf{0.5} & 0.4 & 0.3 &- &- &- &- &\underline{0.03} & \underline{0.03} & \underline{0.03} & 0.03 & 1,377 & 398 & 21 & 7 \cr
			&\pbcpc&5,585 & 5,707 & 3,596 & 1,241 &1 & 0.6 & 0.4 & 0.3 &- &- &- &- &0.03 & 0.03 & 0.03 & 0.03 &1,721 & 592 & 32 & \underline{6}\cr
			&\pbcpcp& 2,327 & 2,878 & 1,991 & 713 &1 & 0.6 & 0.4 & 0.3 &- &- &- &- &0.03 & 0.03 & 0.03 & 0.03 &1,110 & 447 & 29 & 7 \cr
			&\biclique& - & 109,345 & 4,728 & 8,555 &2,315 & 6 & 0.4 & 0.3 & - & - & 401,158 & 23,651 & 0.09 & 0.04 & 0.04 & 0.04 & - & 77 & 20 & 14 \cr
			&\bicliquem& \underline{33} & \underline{117} & \underline{101} & \underline{38} &\underline{1} & 0.6  & \underline{0.4}  & \underline{0.3}  &\underline{5,219} & \underline{146,064} & \underline{103,754} & \underline{2,324} & \textbf{0.03} & \textbf{0.03} & \textbf{0.03} & \underline{0.03} & \underline{30} & \underline{22} & \underline{9} & 7 \cr
			&\bicliquep& \textbf{19} & \textbf{45} & \textbf{33} & \textbf{13} &\textbf{1} & \underline{0.6} & \textbf{0.4} & \textbf{0.3} &\textbf{4,066} & \textbf{61,855} & \textbf{36,416} & \textbf{1,252} &0.04 & 0.04 & 0.04 & \textbf{0.03} &\textbf{30} & \textbf{16} & \textbf{9} & \textbf{6} \cr
			\hline
			\multicolumn{2}{|c|}{Datasets} &  \multicolumn{4}{c|}{\kw{Twitter}} & \multicolumn{4}{c|}{\kw{Imdb}} & \multicolumn{4}{c|}{\kw{Actor2}} & \multicolumn{4}{c|}{\kw{Amazon}}&\multicolumn{4}{c|}{\kw{DBLP}} \cr 
			\hline
			$\alpha$&\diagbox{Algo}{$\beta$}&2&4&6&8&2&4&6&8&2&4&6&8&2&4&6&8&2&4&6&8 \cr
			\hline
			\multirow{6}{*}{2}&\MBAG&11,652 & 5,290 & 1,817 & 657 &859 & 706 & 622 & 409 &7,303 & 6,650 & 5,217 & 3,502 &16,282 & 3,974 & 580 & 137 &\underline{18} & \textbf{12} & \underline{10} & \underline{9}\cr
			&\pbcpc&47,374 & 24,237 & 6,055 & 2,335 &323 & 188 & 293 & 418 &1,820 & 2,285 & 2,377 & 2,041 &20,703 & 6,687 & 1,297 & 342 &28 & 19 & 15 & 13\cr
			&\pbcpcp& 10,580 & 11,843 & 3,162 & 1,337 &102 & 58 & 90 & 160 &266 & 524 & 772 & 780 &2,413 & 3,067 & 706 & 228 &23 & 15 & 12 & 10\cr
			&\biclique& 685 & 5,339 & 4,584 & - &55 & 147 & 58,284 & - & 190 & 875 & 39,764 & - & 964 & 727 & - & - & 20 & 176 & 296,803 & -   \cr
			&\bicliquem&\textbf{510} & \underline{777} & \underline{208} & \underline{108} &\textbf{35} & \textbf{26} & \textbf{23} & \textbf{21} &\underline{88} & \underline{56} & \underline{48} & \underline{47} &\underline{618} & \textbf{157} & \textbf{79} & \textbf{55} &\textbf{17} & \underline{13} & \textbf{10} & \textbf{9}\cr
			&\bicliquep&\underline{518} & \textbf{402} & \textbf{152} & \textbf{94} &\underline{42} & \underline{33} & \underline{28} & \underline{25} &\textbf{79} & \textbf{51} & \textbf{44} & \textbf{39} &\textbf{590} & \underline{176} & \underline{108} & \underline{82} &25 & 19 & 16 & 14\cr
			\hline
			\multirow{6}{*}{4}&\MBAG& 4,339 & 2,737 & 1,236 & 457 &762 & 687 & 614 & 403 &7,107 & 6,586 & 5,178 & 3,468 &7,856 & 1,720 & 227 & 57 &\textbf{7} & 5 & 4 & 3\cr
			&\pbcpc&11,436 & 6,764 & 3,176 & 1,482 &408 & 355 & 446 & 514 &1,622 & 2,251 & 2,462 & 2,125 &9,234 & 2,496 & 427 & 100 &8 & 5 & 4 & 3\cr
			&\pbcpcp& 6,282 & 4,237 & 2,004 & 1,001 &99 & 74 & 121 & 185 &201 & 516 & 794 & 840 &4,219 & 1,508 & 304 & 78 &8 & 5 & 4 & 4\cr
			&\biclique& - & 12,028 & 20,544 & 242,804 & 92 & 1,805 & 1,335 & 25,431 & 626 & 4,996 & 29,168 & 526,141 &534 & 2,687 & 474 & 9,676 & 12 & 7 & 29 & 1,834 \cr
			&\bicliquem& \underline{134} & \underline{129} & \underline{94} & \underline{87} &\textbf{22} & \underline{21} & \underline{26} & \underline{21} &\underline{57} & \underline{131} & \underline{51} & \underline{44} &\underline{303} & \underline{161} & \underline{61} & \textbf{39} &\underline{8} & \underline{5} & \underline{4} & \underline{3}\cr
			&\bicliquep& \textbf{134} & \textbf{100} & \textbf{69} & \textbf{64} &\underline{23} & \textbf{18} & \textbf{16} & \textbf{15} &\textbf{47} & \textbf{48} & \textbf{34} & \textbf{31} &\textbf{280} & \textbf{111} & \textbf{61} & \underline{42} &9 & \textbf{5} & \textbf{4} & \textbf{3}\cr
			\hline
			\multirow{6}{*}{6}&\MBAG& 2,184 & 1,596 & 891 & 298 &623 & 600 & 536 & 340 &6,859 & 6,376 & 5,085 & 3,343 &1,576 & 349 & 63 & \underline{32} &3 & 2 & 2 & 2\cr
			&\pbcpc&3,124 & 2,404 & 1,677 & 690 &557 & 551 & 605 & 568 &1,988 & 2,585 & 2,583 & 2,270 &1,669 & 499 & 97 & 44 &3 & 2 & 2 & 2\cr
			&\pbcpcp& 2,666 & 1,469 & 1,073 & 506 &169 & 164 & 204 & 237 &326 & 641 & 933 & 955 &1,035 & 331 & 71 & 36 &3 & 3 & 2 & 2\cr
			&\biclique& - & - & 134,078 & 27,608 & 1,748 & 835 & 11,021 & 10,968 &194,453 & 22,912 & 49,668 & 137,995 & 33,212 & 1,595 & 525 & 806 & 10,632 & 6 & 17 & 30 \cr
			&\bicliquem& \textbf{90} & \underline{81} & \underline{68} & \underline{52} &\textbf{17} & \underline{17} & \underline{22} & \underline{23} &\underline{51} & \underline{121} & \underline{62} & \underline{52} &\textbf{171} & \underline{80} & \textbf{44} & \textbf{32} &\underline{3} & \underline{2} & \underline{2} & \underline{2}\cr
			&\bicliquep& \underline{95} & \textbf{58} & \textbf{51} & \textbf{41} &\underline{19} & \textbf{14} & \textbf{14} & \textbf{13} &\textbf{40} & \textbf{43} & \textbf{38} & \textbf{32} &\underline{213} & \textbf{79} & \underline{47} & 33 &\textbf{3} & \textbf{2} & \textbf{2} & \textbf{2}\cr
			\hline
			\multirow{6}{*}{8}&\MBAG& 1,482 & 1,166 & 651 & 210 &400 & 388 & 340 & 187 &6,381 & 5,838 & 4,529 & 2,958 &445 & 104 & 41 & \underline{28} &\underline{1} & 1 & 1 & 1\cr
			&\pbcpc&1,479 & 1,297 & 994 & 369 &440 & 437 & 450 & 331 &2,631 & 2,922 & 2,867 & 2,398 &539 & 143 & 54 & 34 &1 & 1 & 1 & 2\cr
			&\pbcpcp& 880 & 729 & 669 & 271 &187 & 186 & 209 & 193 &625 & 894 & 1,093 & 1,023 &338 & 101 & 45 & 31 &2 & 1 & 1 & 1 \cr
			&\biclique&- & - & - & 190,333 & 355,545 & 2,375 & 1,985 & 11,003 &- &- &- &- &- &29,402 & 3,606 & 355 &- & 25 & 61 & 93 \cr
			&\bicliquem& \textbf{74} & \underline{65} & \underline{53} & \underline{36} &\textbf{15} & \underline{13} & \underline{16} & \underline{16} &\underline{46} & \underline{100} & \underline{151} & \underline{66} &\textbf{143} & \textbf{59} & \textbf{38} & \textbf{28} &\textbf{1} & \underline{1} & \underline{1} & \underline{1} \cr
			&\bicliquep& \underline{79} & \textbf{45} & \textbf{36} & \textbf{28} &\underline{16} & \textbf{12} & \textbf{12} & \textbf{11} &\textbf{35} & \textbf{38} & \textbf{46} & \textbf{36} &\underline{188} & \underline{64} & \underline{40} & 29 &2 & \textbf{1} & \textbf{1} & \textbf{1}\cr
			\hline
	\end{tabular}}
	\vspace*{-0.4cm}
\end{table*}

\stitle{Algorithms.} We first implement the state-of-the-art algorithm in \cite{lehmann2008biclique,chen2023index} as \MBAG.

\begin{itemize}
	\item \MBAG (Algorithm~\ref{alg:bcpc_baseline}) is directly based on the maximal biclique adjacency graph (also abbreviated as \MBAG). It first enumerates all maximal bicliques, then constructs the maximal biclique adjacency graph using these bicliques, and finally traverses this adjacency graph to compute the \BCPC.
\end{itemize}

We then implement the following partial-\BCPC based solution and $(\alpha, \beta)$-biclique based solution.

\begin{itemize}
	\item \pbcpc (Algorithm~\ref{alg:bcpc_par_bcpc_basic}) is the basic partial-\BCPC based solution. It first enumerates maximal bicliques, and builds the \MBE tree. For each $(\alpha, \beta)$-node in the \MBE tree, it identifies all maximal bicliques associated with that node and connects them into a single partial-\BCPC. It then traverses the \MBAG reduced by the partial-\BCPCs to obtain the final \BCPC.

	\item \pbcpcp (Algorithm~\ref{alg:bcpc_par_bcpc_slabel}) is an improved version of \pbcpc. The key difference lies in its use of stop-labels to reduce the search overhead when identifying all maximal bicliques associated with an $(\alpha, \beta)$-node.
	
	\item \biclique (Algorithm~\ref{alg:bcpc_ablist_frame}) is the basic solution based on $(\alpha, \beta)$-biclique enumeration. Its core idea is to enumerate all $(\alpha, \beta)$-bicliques, and for each such biclique, connect all maximal bicliques that contain it in order to construct the \BCPC.
	
	\item \bicliquem (Algorithm~\ref{alg:bcpc_ablist_mbiclique}) is an improved version of \biclique. It prunes $(\alpha, \beta)$-biclique enumeration branches by checking whether the maximal bicliques associated with the currently enumerated biclique already belong to the same \BCPC.
	
	\item \bicliquep (Algorithm~\ref{alg:bcpc_ablist_pbcpc}) is also an improved version of \biclique. Unlike \bicliquem, it leverages the method from \pbcpcp to compute partial-\BCPCs in advance, thereby identifying some maximal bicliques that already belong to the same \BCPC. This enables a stronger pruning effect.
	
\end{itemize}

\stitle{Parameters.} There are two parameters used in the experiments: $\alpha$ and $\beta$. Their default values are both set to 4, and the possible values they can take are 2, 4, 6, and 8.

All the algorithms are implemented in C++. The experiments are conducted on a system running Ubuntu 20.04.4 LTS with an AMD Ryzen 3990X 2.2GHz CPU and 256GB of memory.

\subsection{Experimental Results}
\stitle{Exp-1: Efficiency of various algorithms.} In this experiment, we evaluated the performance of all algorithms across all datasets, and the results are presented in Table~\ref{tab:all_performance}.

Overall, \MBAG and \pbcpc algorithms perform the worst in terms of runtime across most datasets and $ \alpha,\beta $. This is mainly because the \MBAG algorithm traverses the original maximal biclique adjacency graph (\MBAG), which is often very large (see Table~\ref{tab:datasets}). Although the \pbcpc algorithm attempts to reduce the size of the \MBAG by computing partial-\BCPCs, these partial-\BCPCs are not sufficiently comprehensive, leading to limited pruning effectiveness on the \MBAG (we will explore the effectiveness of these partial-\BCPCs in Exp-3).

\pbcpcp performs slightly better than \MBAG and \pbcpc, especially when $\alpha$ and $ \beta $ are small. For instance, on dataset \kw{Youtube}, when $ \alpha=\beta=2 $, \pbcpcp outperforms \MBAG and \pbcpc by more than one order of magnitude. This is because when $\alpha$ and $ \beta $ are small, there is a larger number of maximal bicliques that are larger than $ (\alpha, \beta) $-bicliques, which increases the scale of maximal biclique adjacency graph. In such cases, the effect of \pbcpcp using partial-\BCPC to reduce the size of the adjacency graph is more prominently demonstrated.

\biclique achieves relatively good performance only when both $\alpha$ and $ \beta $ are small—for example, on datasets \kw{Youtube}, \kw{Bookcrossing}, and \kw{DBLP} with $ \alpha=\beta=2 $. When $\alpha$ or $ \beta $ increases, the performance of \biclique drops sharply. This is because the number of $ (\alpha,\beta) $-bicliques grows exponentially as $\alpha$ and $ \beta $ increase.

\bicliquem and \bicliquep are the two optimal algorithms in almost all cases, and they can outperform \MBAG by nearly three orders of magnitude. For example, on dataset \kw{Youtube}, when $ \alpha=\beta=2 $, \bicliquep only takes 34 seconds, while \MBAG requires 13,376 seconds. Furthermore, \bicliquep performs better than \bicliquem, and this advantage is more pronounced on dataset \kw{Github} (under many parameters, only \bicliquep can complete within 7 days, i.e., 604,800 seconds). This is because \bicliquep uses partial-\BCPC for pruning, which has stronger pruning capability compared to maximal biclique (see Exp-4).

\begin{figure}[t]
	\begin{center}
		\begin{tabular}[t]{c}
			\hspace*{-0cm}
			\subfigure[{\scriptsize \kw{Youtube}, $ \alpha=2 $, varying $ \beta $}]{
				\includegraphics[width=0.44\columnwidth]{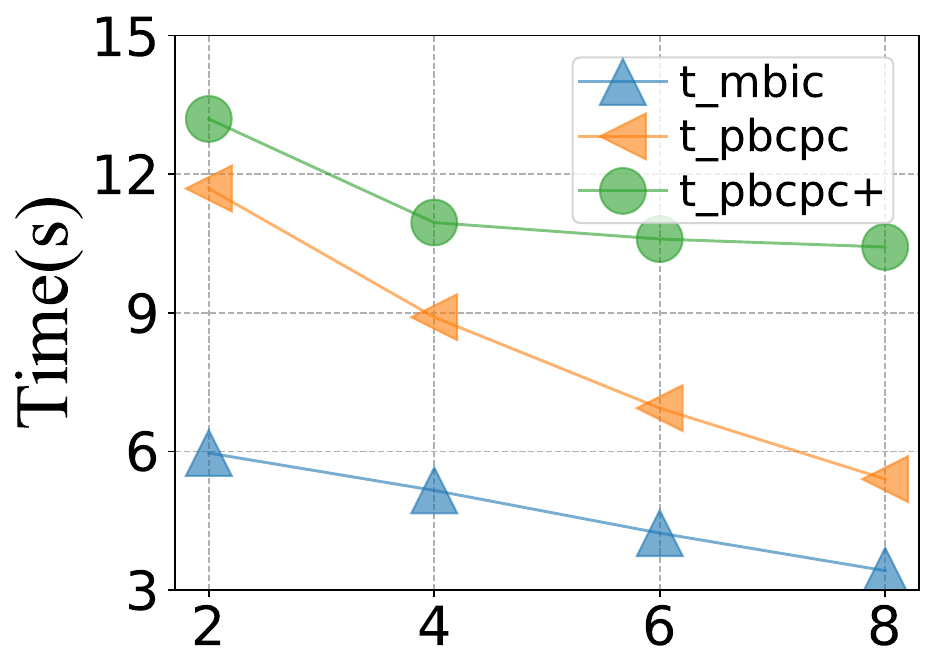}
			}
			\hspace*{-0cm}
			\subfigure[{\scriptsize \kw{Youtube}, $ \alpha=8 $, varying $ \beta $}]{
				\includegraphics[width=0.42\columnwidth]{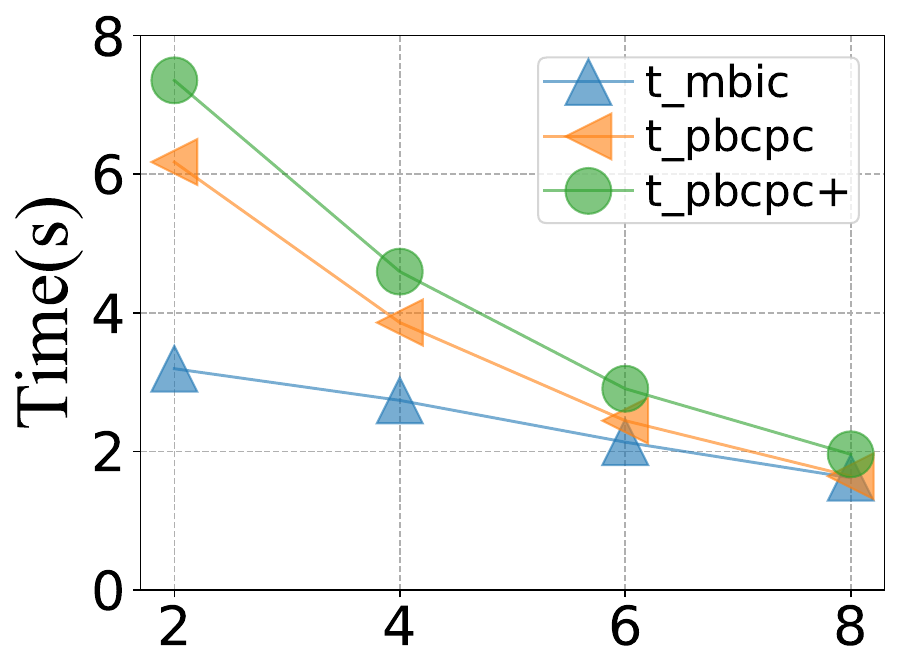}
			}
			\\
			\hspace*{-0cm}
			\subfigure[{\scriptsize \kw{Github}, $ \alpha=2 $, varying $ \beta $}]{
				\includegraphics[width=0.44\columnwidth]{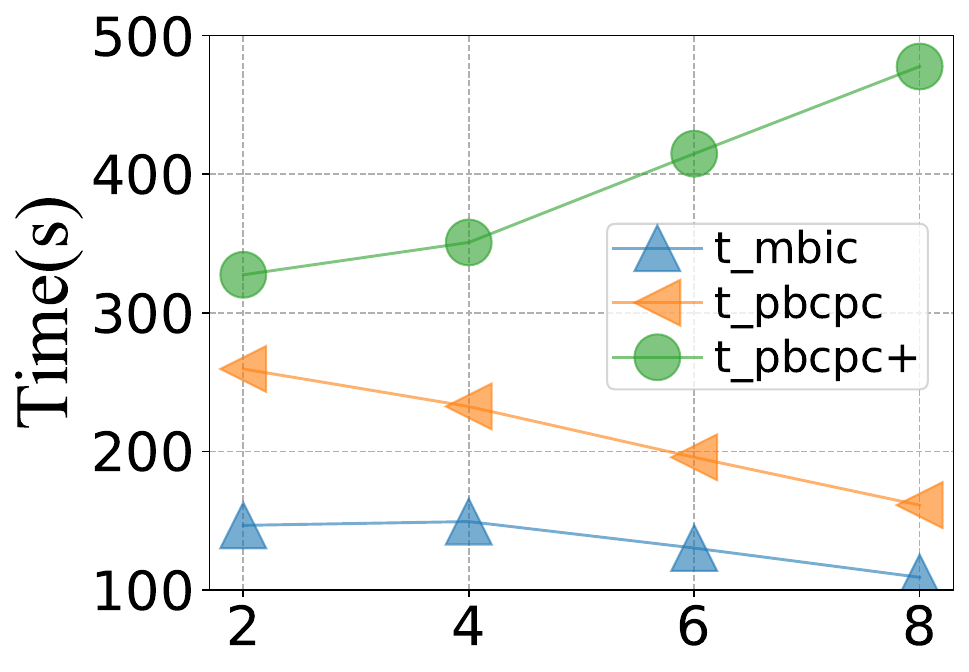}
			}
			\hspace*{-0cm}
			\subfigure[{\scriptsize \kw{Github}, $ \alpha=8 $, varying $ \beta $}]{
				\includegraphics[width=0.44\columnwidth]{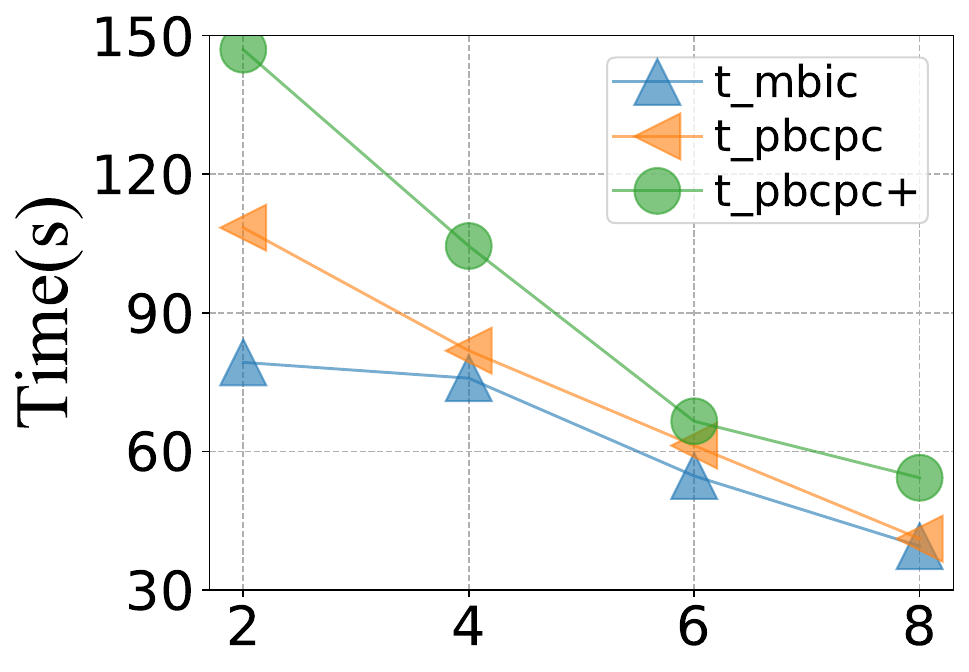}
			}
			\\
			\hspace*{-0cm}
			\subfigure[{\scriptsize \kw{Twitter}, $ \alpha=2 $, varying $ \beta $}]{
				\includegraphics[width=0.44\columnwidth]{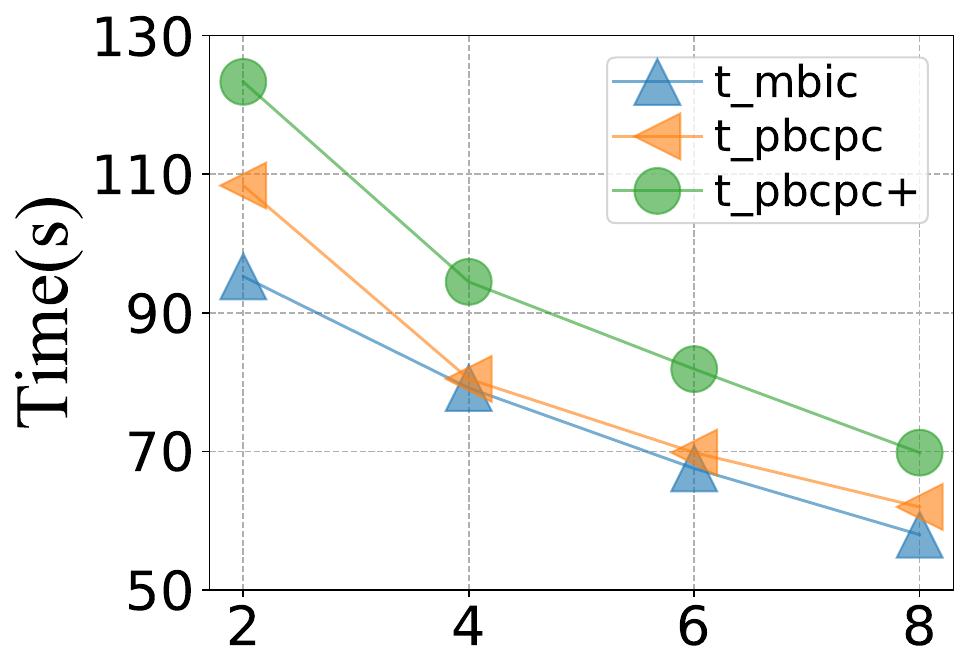}
			}
			\hspace*{-0cm}
			\subfigure[{\scriptsize \kw{Twitter}, $ \alpha=8 $, varying $ \beta $}]{
				\includegraphics[width=0.42\columnwidth]{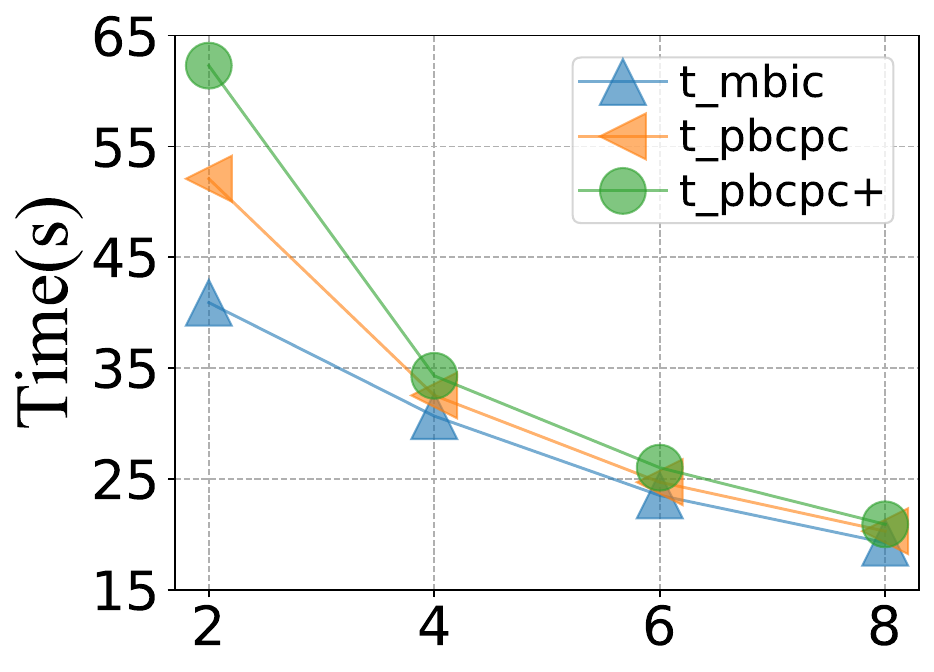}
			}
		\end{tabular}
	\end{center}
	\vspace*{-0.3cm}
	\caption{Time spent on maximal biclique enumeration (\kw{t\_mbic}), partial-\BCPC computation (including maximal biclique enumeration as a necessary step) in \pbcpc (\kw{t\_pbcpc}) and \pbcpcp (\kw{t\_pbcpc+})}
	\vspace*{-0.4cm}
	\label{fig:pbcpc_time}
\end{figure}

\stitle{Exp-2: Overhead of computing partial-\BCPC.} This experiment evaluates the overhead of computing partial-\BCPCs (with maximal biclique enumeration included, which is a necessary step). Specifically, we compare the time spent on computing partial-\BCPCs in the algorithms \pbcpc and \pbcpcp, and also compare it against the time spent on maximal biclique enumeration. The experimental results are shown in Figure~\ref{fig:pbcpc_time}.

As shown in Figure~\ref{fig:pbcpc_time}, although computing partial-\BCPCs introduces additional overhead on top of maximal biclique enumeration, the added cost is manageable. For example, on datasets \kw{Youtube} and \kw{Twitter}, \kw{t\_pbcpc} and \kw{t\_pbcpc+} are only slightly higher than the time spent on maximal biclique enumeration alone (\kw{t\_mbic}). For the \kw{Github} dataset, due to the larger number of maximal bicliques (see Table~\ref{tab:datasets}) and larger scale of the \MBE tree, \kw{t\_pbcpc+} is no longer close to \kw{t\_mbic} and \kw{t\_pbcpc}. However, this overhead is meaningful—our best-performing algorithm, \bicliquep, is built upon the partial-\BCPC computed in \pbcpcp (the time spent is also \kw{t\_pbcpc+}). Notably, \bicliquep is the only algorithm that successfully passed all tests on the \kw{Github} dataset.

\begin{figure}[t]
	\begin{center}
		\begin{tabular}[t]{c}
			\hspace*{-0cm}
			\subfigure[{\scriptsize \kw{Youtube}, $ \alpha=2 $, varying $ \beta $}]{
				\includegraphics[width=0.44\columnwidth]{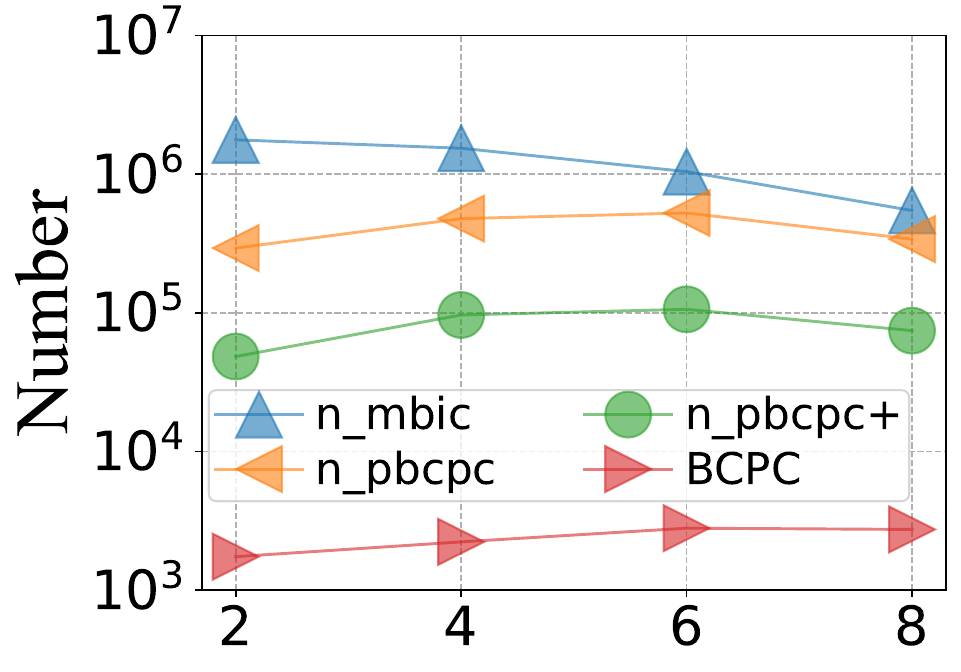}
			}
			\hspace*{-0cm}
			\subfigure[{\scriptsize \kw{Youtube}, $ \alpha=8 $, varying $ \beta $}]{
				\includegraphics[width=0.44\columnwidth]{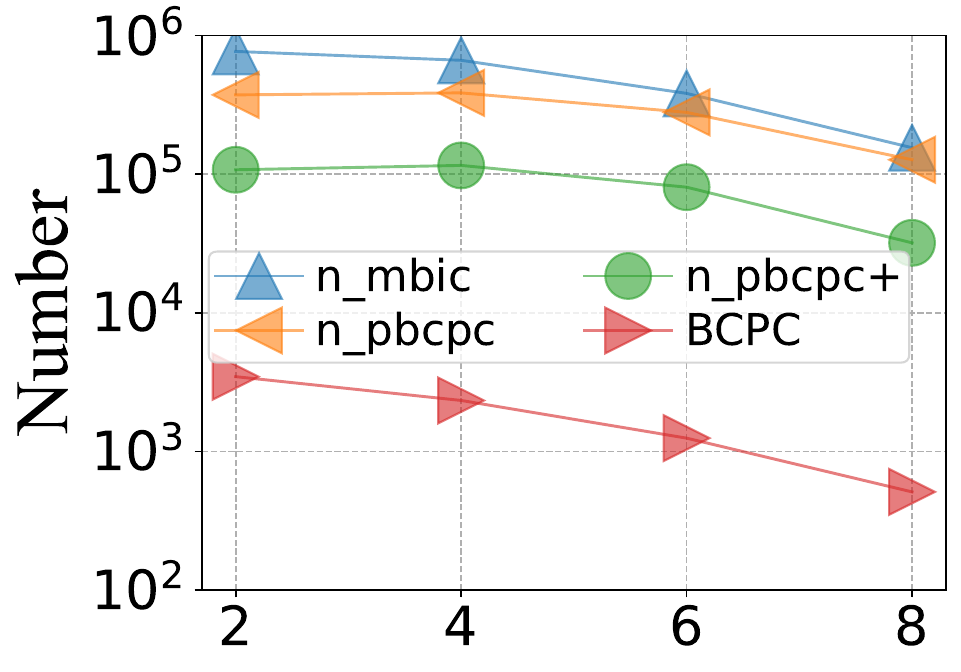}
			}
			\\
			\hspace*{-0cm}
			\subfigure[{\scriptsize \kw{Github}, $ \alpha=2 $, varying $ \beta $}]{
				\includegraphics[width=0.44\columnwidth]{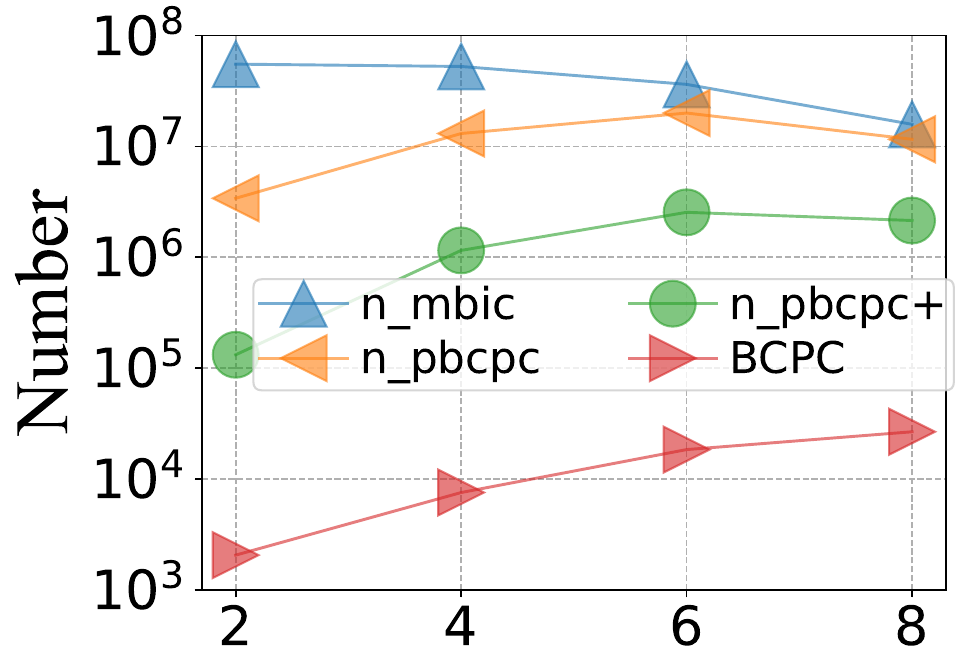}
			}
			\hspace*{-0cm}
			\subfigure[{\scriptsize \kw{Github}, $ \alpha=8 $, varying $ \beta $}]{
				\includegraphics[width=0.44\columnwidth]{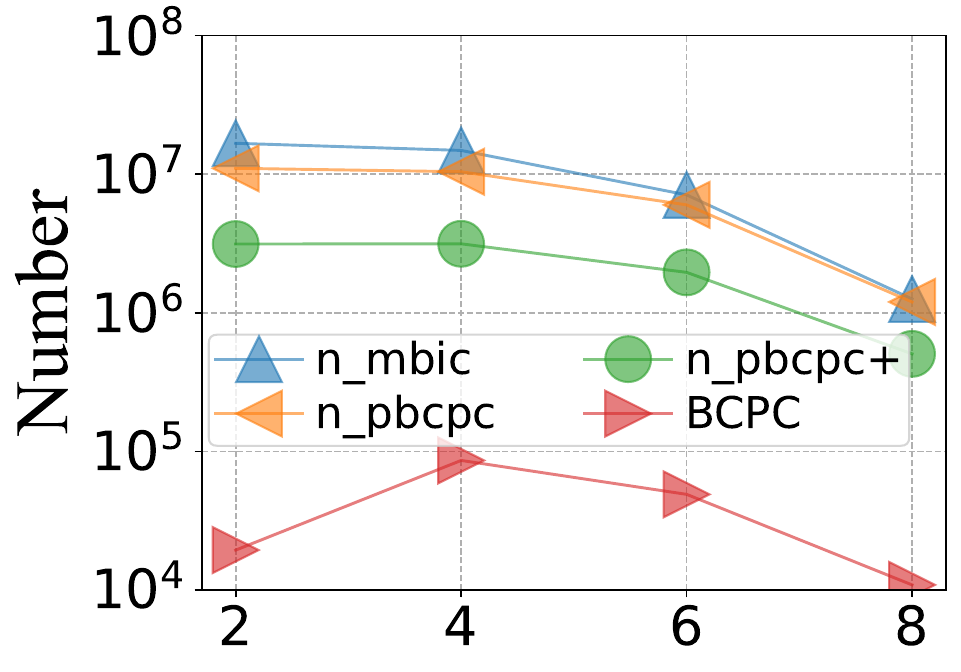}
			}
			\\
			\hspace*{-0cm}
			\subfigure[{\scriptsize \kw{Twitter}, $ \alpha=2 $, varying $ \beta $}]{
				\includegraphics[width=0.44\columnwidth]{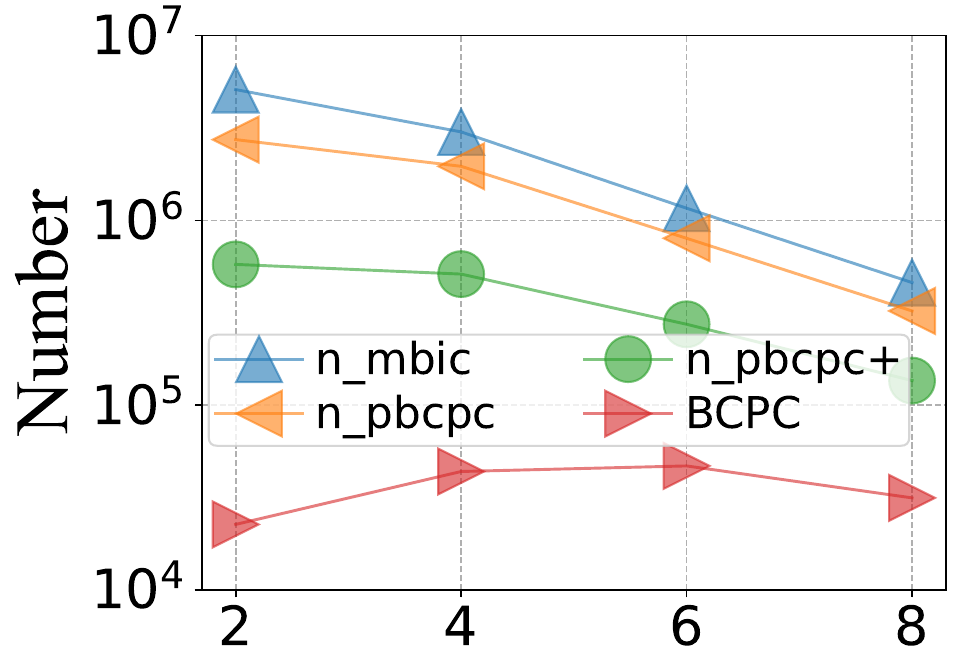}
			}
			\hspace*{-0cm}
			\subfigure[{\scriptsize \kw{Twitter}, $ \alpha=8 $, varying $ \beta $}]{
				\includegraphics[width=0.44\columnwidth]{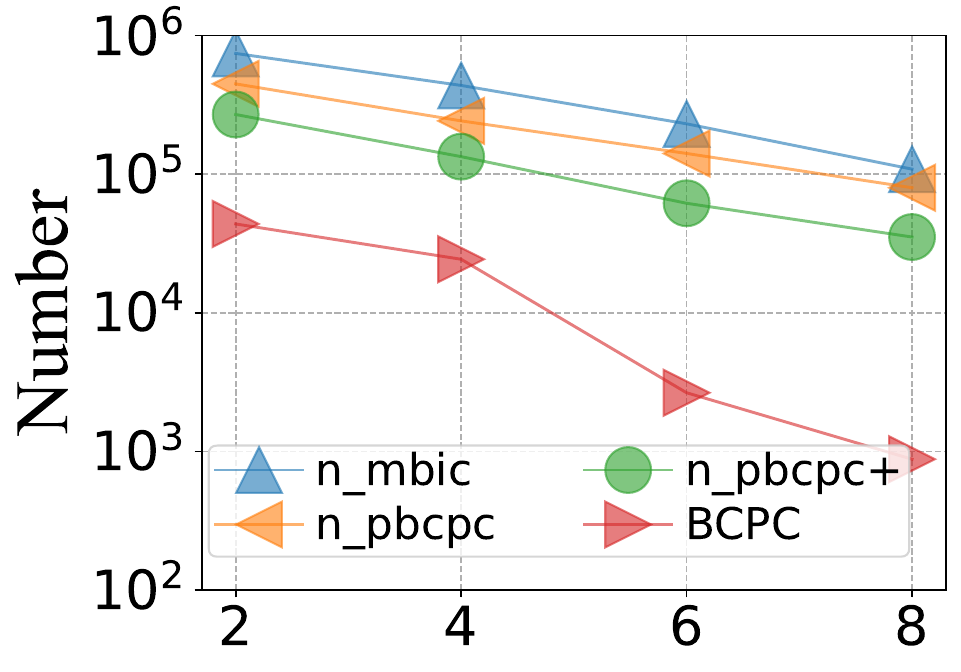}
			}
		\end{tabular}
	\end{center}
	\vspace*{-0.3cm}
	\caption{Number of maximal bicliques (\kw{n\_mbic}, where the size of $ X $ set and $ Y $ set is at least $ \alpha $ and $ \beta $ respectively), partial-\BCPCs obtained by \pbcpc (\kw{n\_pbcpc}) and \pbcpcp (\kw{n\_pbcpc+}), and number of \BCPCs}
	\vspace*{-0.3cm}
	\label{fig:pbcpc_num}
\end{figure}

\stitle{Exp-3: Effectiveness of partial-\BCPC in partial-\BCPC based solution.} This experiment investigates the effectiveness of partial-\BCPC in \pbcpc and \pbcpcp. Specifically, we compare the number of partial-\BCPCs obtained by the two algorithms with the number of maximal bicliques and the number of \BCPCs. The results are shown in Figure~\ref{fig:pbcpc_num}. We can see that in all cases, the inequality in Theorem~\ref{thm:pbcpc_basic_num} and~\ref{thm:pbcpc_num_op} holds: \kw{n\_mbic} $ \geq $ \kw{n\_pbcpc} $ \geq $ \kw{n\_pbcpc+} $ \geq $ \BCPC. We can also observe that compared to \kw{n\_mbic} and \kw{n\_pbcpc}, \kw{n\_pbcpc+} is much closer to \BCPC, and in some cases, it is more than an order of magnitude smaller than \kw{n\_mbic} (see Figure~\ref{fig:pbcpc_num} (a), (c)). In both the \pbcpc and \pbcpcp algorithms, vertices in the \MBAG are regrouped according to the partial-\BCPCs, meaning that the number of vertices in the reduced \MBAG equals the number of partial-\BCPCs. As a result, \pbcpcp outperforms \pbcpc on nearly all datasets.

\begin{figure}[t]
	\begin{center}
		\begin{tabular}[t]{c}
			\hspace*{-0cm}
			\subfigure[{\scriptsize \kw{Youtube}, $ \alpha=2 $, varying $ \beta $}]{
				\includegraphics[width=0.44\columnwidth]{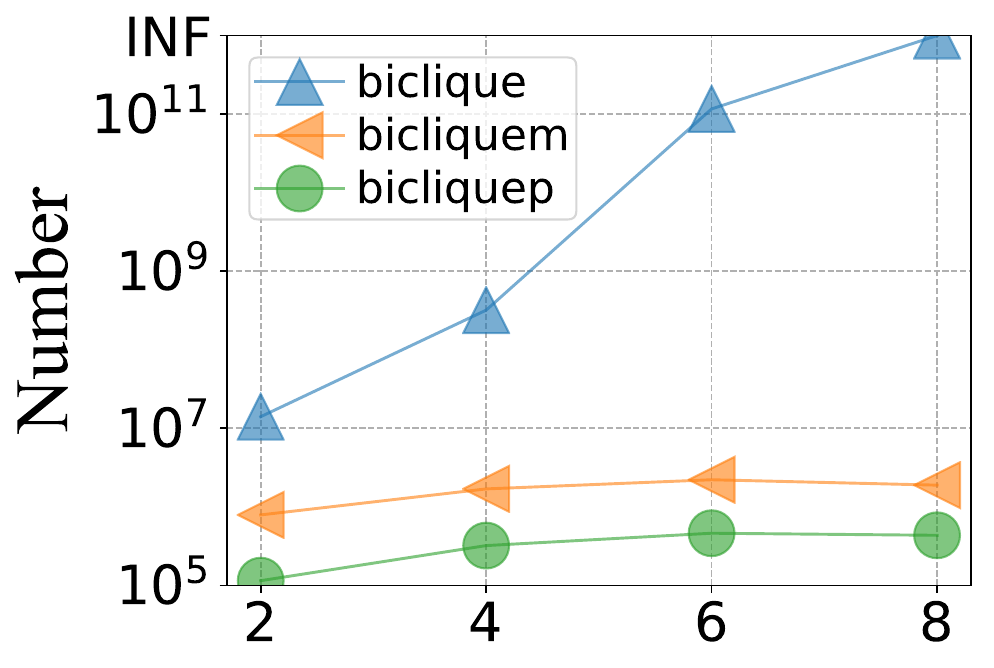}
			}
			\hspace*{-0cm}
			\subfigure[{\scriptsize \kw{Youtube}, $ \alpha=8 $, varying $ \beta $}]{
				\includegraphics[width=0.44\columnwidth]{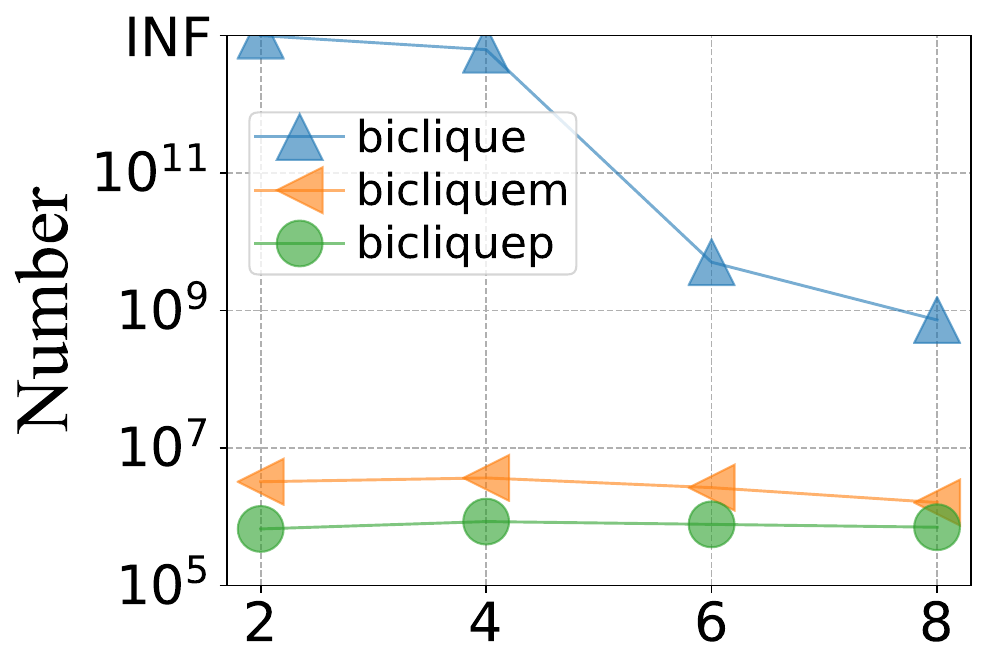}
			}
			\\
			\hspace*{-0cm}
			\subfigure[{\scriptsize \kw{Github}, $ \alpha=2 $, varying $ \beta $}]{
				\includegraphics[width=0.44\columnwidth]{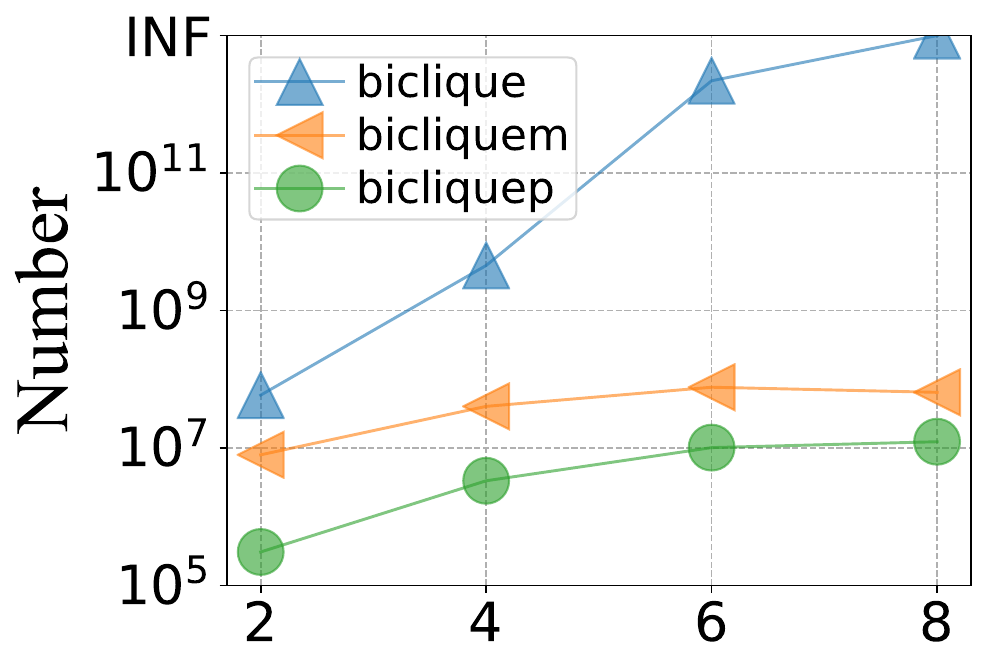}
			}
			\hspace*{-0cm}
			\subfigure[{\scriptsize \kw{Github}, $ \alpha=8 $, varying $ \beta $}]{
				\includegraphics[width=0.44\columnwidth]{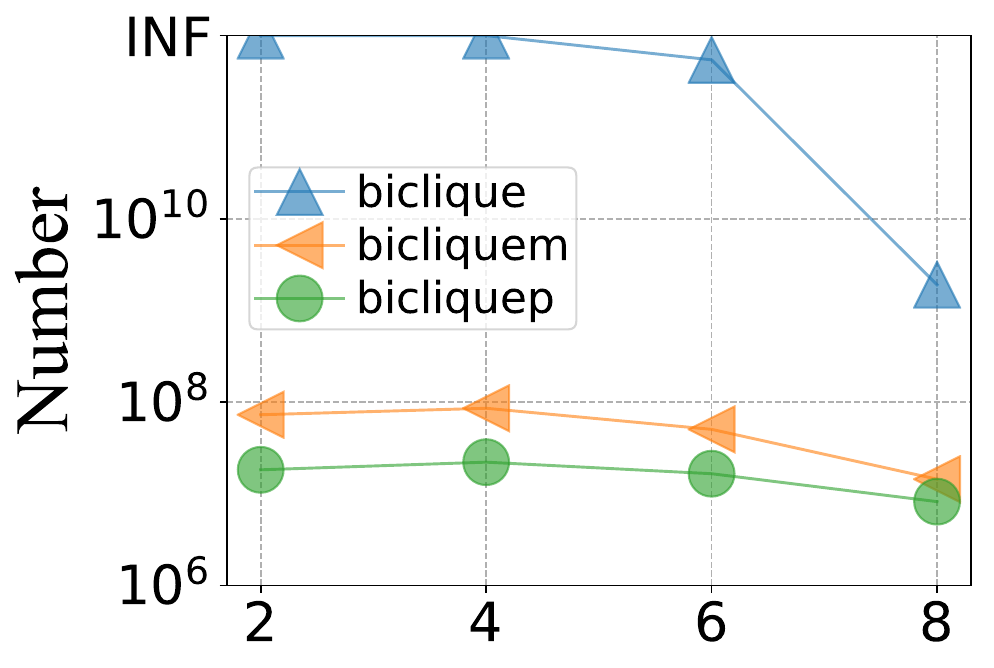}
			}
			\\
			\hspace*{-0cm}
			\subfigure[{\scriptsize \kw{Twitter}, $ \alpha=2 $, varying $ \beta $}]{
				\includegraphics[width=0.44\columnwidth]{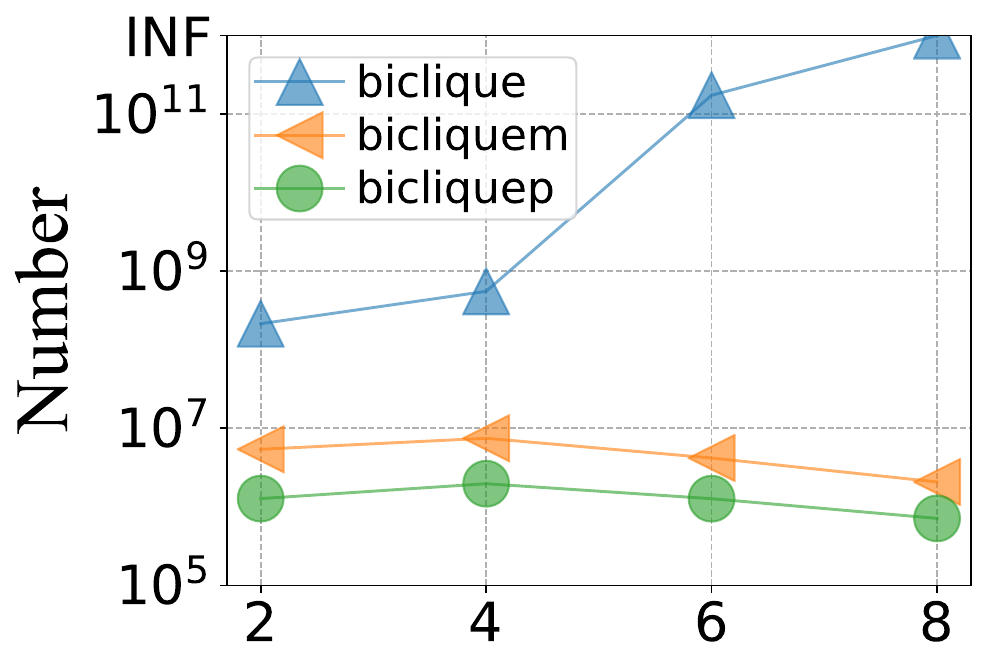}
			}
			\hspace*{-0cm}
			\subfigure[{\scriptsize \kw{Twitter}, $ \alpha=8 $, varying $ \beta $}]{
				\includegraphics[width=0.44\columnwidth]{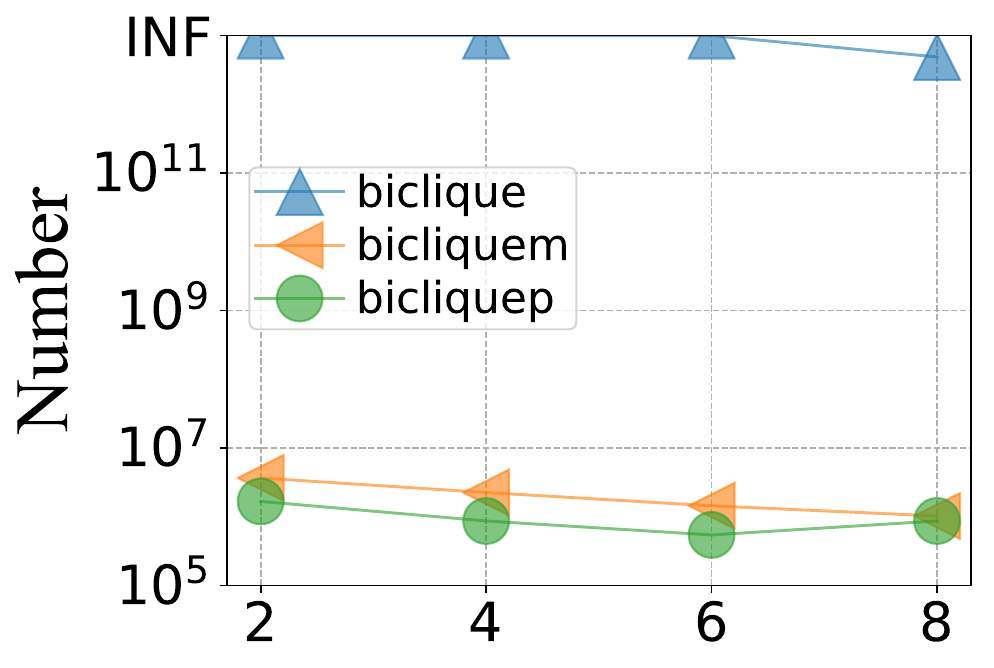}
			}
		\end{tabular}
	\end{center}
	\vspace*{-0.3cm}
	\caption{Number of enumerated nodes in the $ (\alpha, \beta) $-biclique enumeration trees of three algorithms: \biclique (\kw{biclique}), \bicliquem (\kw{bicliquem}), and \bicliquep (\kw{bicliquep})}
	\vspace*{-0.4cm}
	\label{fig:qpbcl_node}
\end{figure}

\stitle{Exp-4: Effectiveness of partial-\BCPC in $ (\alpha,\beta) $-biclique based solution.} This experiment investigates the pruning effectiveness of maximal biclique and partial-\BCPC in the $ (\alpha, \beta) $-biclique enumeration process. Figure~\ref{fig:qpbcl_node} shows the number of enumerated nodes in the $ (\alpha, \beta) $-biclique enumeration trees of three algorithms: \biclique, \bicliquem, and \bicliquep. It can be observed that in \biclique (without any pruning), the number of nodes in the enumeration tree is more than four orders of magnitude higher than that in the other two algorithms, which sufficiently demonstrates the pruning capabilities of maximal biclique and partial-\BCPC in $ (\alpha, \beta) $-biclique enumeration process. Among \bicliquem and \bicliquep, the partial-\BCPC in \bicliquep exhibits the strongest pruning capability, as the number of enumerated nodes in \bicliquep is another order of magnitude smaller than that in \bicliquem (e.g., when $ \beta=2 $ in Figure~\ref{fig:qpbcl_node} (c)). This explains why \bicliquep achieves the best performance in most cases in Table~\ref{tab:all_performance}, particularly with the dataset \kw{Github}.

\stitle{Exp-5: Scalability.} This experiment explores the scalability of our best algorithm, \bicliquep. Table~\ref{tab:scala} presents the performance of \bicliquep on \kw{Github} and four other large datasets. Specifically, we sampled 20\%, 40\%, 60\%, and 80\% of the edges from each dataset respectively. It can be observed that as the dataset size increases, the growth of running time of \bicliquep is manageable, demonstrating that \bicliquep has excellent scalability.

\begin{table}[t!]
	\small
	\centering
	\caption{Scalability test of \pbcpcp (s)} \vspace*{-0.2cm} \label{tab:scala}
	\resizebox*{0.8\columnwidth}{!}{\begin{tabular}{|c | c c c c c |}
			\hline
			\rule{0pt}{8pt}&\kw{Github}&\kw{Imdb}&\kw{Actor2} & \kw{Amazon} &\kw{DBLP}   \cr
			\hline
			\rule{0pt}{7pt}20\%&5,779.7  & 3.4  & 1.2  & 8.3  & 0.8 \cr
			40\%&72,202.2  & 7.8  & 5.7  & 54.5  & 1.4 \cr
			60\%&119,608.0  & 13.5  & 14.7  & 85.2  & 2.8 \cr
			80\%&142,445.8  & 17.5  & 25.8  & 107.3  & 4.5 \cr
			100\%&145,121.7  & 18.2  & 49.0  & 111.2  & 5.8 \cr
			\hline
	\end{tabular}}
	\vspace*{-0.2cm}
\end{table}

\stitle{Exp-6: Memory.} Table~\ref{tab:mem} presents the memory usage of all algorithms on the largest four datasets and dataset \kw{Github}, which contains the maximum number of maximal bicliques. It can be observed that across all datasets, the three algorithms—\MBAG, \biclique, and \bicliquem—consistently consume the least memory. This is because these algorithms do not require storing the \MBE tree. However, although our best algorithm, \bicliquep, also stores the \MBE tree, its memory consumption remains manageable.

\begin{table}[t!]
	\small
	\centering
	\caption{Memory usage (MB) of various algorithms} \vspace*{-0.2cm} \label{tab:mem}
	\resizebox*{0.95\columnwidth}{!}{\begin{tabular}{|c | c c c c c c|}
			\hline
			\rule{0pt}{8pt}&\MBAG&\pbcpc&\pbcpcp & \biclique &\bicliquem &\bicliquep   \cr
			\hline
			\rule{0pt}{7pt}\kw{Github}& 10,051 & 26,965 & 26,418 & 9,904 & \textbf{9,904} & 22,158 \cr
			\kw{Imdb} &\textbf{233} & 364 & 361 & 260 & 260 & 343 \cr
			\kw{Actor2}&\textbf{409} & 725 & 720 & 449 & 449 & 639  \cr
			\kw{Amazon}&\textbf{711} & 1,228 & 1,207 & 797 & 772 & 1,130 \cr
			\kw{DBLP} &935 & 957 & 970 & 1,200 & \textbf{935} & 940  \cr
			\hline
	\end{tabular}}
	\vspace*{-0.2cm}
\end{table}

\stitle{Exp-7: Case study.} We conduct a case study on the \kw{DBLP} dataset. This dataset includes papers from major journals (TODS, TOIS, TKDE, VLDBJ) and conferences (SIGMOD, SIGKDD, ICDE, SIGIR, VLDB) in the database field, where vertices represent papers and authors respectively, and an edge \((u, v)\) indicates that \(v\) is one of the authors of \(u\). We compare $ (\alpha, \beta) $-\BCPC with maximum biclique, $ k $-bitruss \cite{wang2020efficient}, and $ (\alpha, \beta) $-core \cite{liu2020efficient}.

Given a bipartite graph $ G=(U,V,E) $, $ (\alpha, \beta) $-core of $ G $ is a maximal subgraph of $ G $, $ G_1=(U_1,V_1,E_1) $, that $ \forall u\in U_1,|Nei_{G'}(u)|\geq \alpha,\forall v\in V_1,|Nei_{G'}(v)|\geq \beta $. $ k $-bitruss is a maximal subgraph of $ G $, $ G_2=(U_2,V_2,E_2) $, that each edge in $ E_2 $ is contained in at least $ k $ butterflies (butterfly is a subgraph with four vertices: $ u,w\in U_2,v,x\in V_2 $, $ (u,v)\in E_2,(u,x)\in E_2,(w,v)\in E_2,(w,x)\in E_2 $).

In this case study, $ \alpha=4,\beta=4,k=12 $. By Definition~\ref{def:bcpc}, the smallest maximal bicliques in $ (4, 4) $-\BCPC are either $ (5, 4) $-bicliques or $ (4, 5) $-bicliques, and the edges within these maximal bicliques exist in at least \(3 \times 4 = 12\) butterflies, thus, we set $ k=12 $ for $ k $-bitruss.

\begin{figure}[t!]
	\begin{center}
		\includegraphics[width=0.99\columnwidth]{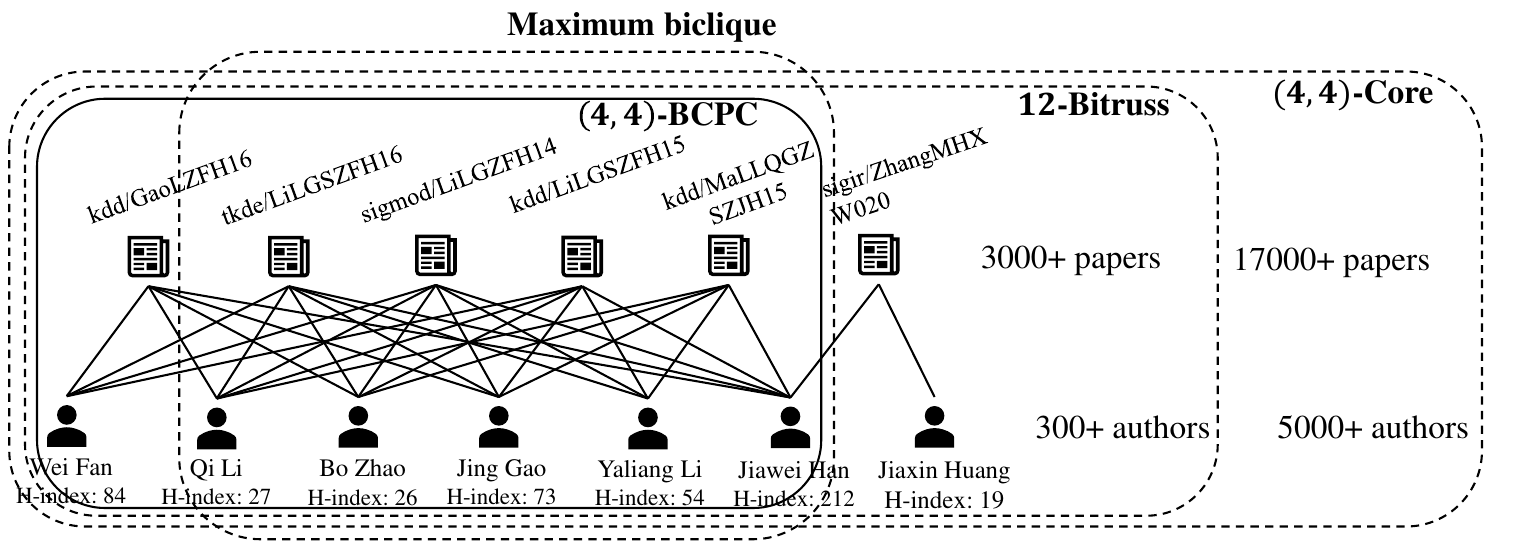}
	\end{center}
	\vspace*{-0.3cm}
	\caption{Case study on \kw{DBLP}}
	\vspace*{-0.5cm}
	\label{fig:case}
\end{figure}

Figure~\ref{fig:case} presents the results containing \textit{Jiawei Han} across all models. Among them, the result from maximum biclique is the smallest. Compared with $ (4, 4) $-\BCPC, it lacks one author, \textit{Wei Fan}, who is one of \textit{Jiawei Han}'s collaborators on Google Scholar. Although $ 12 $-bitruss and $ (4, 4) $-core include the results of $ (4, 4) $-\BCPC, they also contain a large number of other vertices, making the usability of these two models far inferior to that of $ (4, 4) $-\BCPC.

\section{Related Work}
The research related to \BCPC is the most relevant to our work. Initially, \cite{lehmann2008biclique} proposed the \BCPC model and introduced a method to detect \BCPCs by traversing the maximal biclique adjacency graph (\MBAG). Subsequently, \cite{wang2018bmtk} developed an online web toolkit, implementing the approach presented in \cite{lehmann2008biclique}. In the context of social network analysis, \cite{hecking2015analysis} applied \BCPC to identify clusters of users and entities. Regarding Wikipedia networks, \cite{jesus2009bipartite} leveraged \BCPC to explore the relationships between editors and articles, revealing that topics inherently aggregate editors. \cite{liao2010visualizing} applied \BCPC to enterprise networks to analyze interaction patterns between hosts and users, thereby supporting the detection of abnormal user behaviors. Meanwhile, \cite{hecking2014analysis,ziebarth2015resource} utilized \BCPC to study the relationships between students and online learning resources, delving into the temporal evolution patterns between students and courses. Additionally, \cite{chen2023index} investigated the problem of personalized \BCPC search based on the \BCPC model. 

Our work is also related to maximal biclique enumeration and $ (p, q) $-biclique enumeration. Existing methods for maximal biclique enumeration, a foundational task supporting \BCPC model implementation, are primarily rooted in the idea of enumerating the powerset of vertex set \cite{zhang2014finding}. Building on this framework, \cite{abidi2020pivot} introduced the pivot pruning strategy, which effectively reduces redundant search space and improves the efficiency. To further optimize performance, \cite{chen2022efficient} proposed a method that optimizes the vertex enumeration order, enhancing computational speed. Additionally, \cite{dai2023hereditary} developed a unified enumeration framework suitable for hereditary cohesive subgraphs. This framework achieves state-of-the-art performance in maximal biclique enumeration problem, providing a more efficient and generalizable solution for related tasks. Key works on $ (p, q) $-biclique enumeration include \cite{mitzenmacher2015scalable,yang2023p}. \cite{mitzenmacher2015scalable} uses $ (p, q) $-biclique enumeration only as a sub-process for broader graph analysis, while \cite{yang2023p} focuses specifically on optimizing this enumeration task. Additionally, studies like \cite{ye2023efficient} address the exact and approximate counting of $ (p, q) $-bicliques.

There are also several cohesive subgraph models relevant to our work, such as ($\alpha$, $\beta$)-core and $ k $-bitruss. For ($\alpha$, $\beta$)-core, key works include \cite{liu2019efficient,liu2020efficient,liu2023distributed}, which focus on efficient computation and distributed implementation of this structure. In terms of $ k $-bitruss, studies such as \cite{zou2016bitruss,wang2020efficient,lin2025efficient} contribute to optimizing its decomposition and analysis processes. For maximum biclique, \cite{lyu2020maximum,wang2022efficient} propose efficient methods for its identification. Beyond specific subgraph structures, modularity-based modeling approaches (e.g., \cite{barber2007modularity,arthur2020modularity}) and label propagation methods (e.g., \cite{liu2010efficient}) are also studied.

\section{Conclusion}\label{sec:conclusion}

In this work, we tackle the problem of biclique percolation community (\BCPC) detection in bipartite network. Confronting the high time complexity of existing \BCPC detection methods, which stems from the potentially massive maximal biclique adjacency graph (\MBAG), we introduce two innovative solutions. First, the partial-\BCPC based solution reduces redundant computations by grouping maximal bicliques into incomplete \BCPCs, lessening the burden of explicitly representing the entire \MBAG. Second, our novel method based on \((\alpha, \beta)\)-biclique enumeration detects \BCPCs by enumerating all \((\alpha, \beta)\)-bicliques and connecting maximal bicliques that share such bicliques. Leveraging partial-\BCPC, this approach significantly prunes the \((\alpha, \beta)\)-biclique enumeration space. Experimental results demonstrate that our methods achieve remarkable performance, outperforming existing approaches by nearly three orders of magnitude. These advancements not only enhance the efficiency of \BCPC detection but also pave the way for more scalable analysis of bipartite networks in diverse domains, including social networks and recommendation systems.


\bibliographystyle{ACM-Reference-Format}
\balance
\bibliography{bcpc}
\clearpage


\end{document}